\documentclass[12pt,preprint]{aastex}

\begin{document}

\parindent=1.0cm

\title{Red Variable Stars in Three M81 Group Dwarf Galaxies}

\author{T.J. Davidge}

\affil{Dominion Astrophysical Observatory,
\\Herzberg Astronomy \& Astrophysics Research Center,
\\National Research Council of Canada, 5071 West Saanich Road,
\\Victoria, BC Canada V9E 2E7\\tim.davidge@nrc.ca; tdavidge1450@gmail.com}

\begin{abstract}

	Archival [3.6] and [4.5] images are used to identify and characterize 
variable stars in the Magellanic-type galaxies Holmberg II, NGC 2366, 
and IC 2574. Using parametric and non-parametric detection 
methods, 74 confirmed or suspected long period variables (LPVs) are found. 
The period distributions of the LPVs in NGC 2366 and IC 2574 
are similar. While the period distribution of LPVs in Ho II is uncertain due 
to small number statistics there appears to be a deficiency of LPVs with 
periods between 550 and 650 days when compared with NGC 2366 and IC 2574. The LPVs 
are diffusely distributed on the sky, and do not follow the underlying light from 
unresolved stars, as expected if episodes of star formation within the past few hundred 
Myr have occured throughout the galaxies, including their outer regions. 
Distances computed for Ho II and NGC 2366 from the period-luminosity relations 
(PLRs) agree to within $\sim 0.1$ magnitudes with those based on the 
tip of the red giant branch (RGB). Efforts to estimate an LPV-based distance 
modulus for IC 2574 are complicated by the presence of first 
overtone pulsators among LPVs with periods $\sim 600$ days, although the PLR at the long 
period end is consistent with the distance estimated from the RGB-tip. 
In addition to the LPVs, 10 candidate sgB[e] or luminous blue 
variables and 2 candidate red supergiant variables are also identified. Nine candidate 
sgB[e] stars that do not show evidence of variability are also identified based 
on their locations in the color-magnitude diagram.
\end{abstract}

\section{INTRODUCTION}

	The intrinsic high luminosity of long period variables (LPVs) means that they can 
be observed at cosmologically interesting distances, even with telescopes of moderate 
aperture. As LPVs are among the coolest and most luminous stars 
in a galaxy, the contrast with the underlying main body of stars is enhanced in the 
infrared (IR), making it a natural wavelength regime for their study. However, there 
are additional motivations for observing LPVs in the IR. Light in the IR is less prone 
to the molecular line blanketing that depresses the brightnesses of cool stars at shorter 
wavelengths (e.g. Bica et al. 1991), further facilitating the detection of LPVs 
in environments where crowding may otherwise complicate efforts to resolve them. 
LPVs are also standard candles in the IR (e.g Glass \& Lloyd Evans 1981), and 
their intrinsic IR period $vs$ luminosity relation (PLR) is not sensitive to 
galaxy metallicity (Goldman et al. 2019). Unlike many other variable stars, the 
light curves of LPVs have large amplitudes in the IR (e.g. Le Bertre 1993; 
Glass et al. 2001), simplifying their detection. An added bonus 
is that surveys for LPVs in the IR have the potential also to identify highly 
evolved, massive variable stars that are embedded in circumstellar envelopes. The study 
of such stars is of obvious interest for examining the end points of stellar evolution 
at high masses.

	In the current paper, archival images taken with the {\it Spitzer} telescope 
(Werner et al. 2004) are used to search for cool variables in the M81 group 
Magellanic galaxies Holmberg II (Ho II), NGC 2366, and IC 2574. These galaxies have 
similar morphologies, integrated brightnesses, and distances. 
Cignoni et al. (2019) find that dwarf irregular galaxies such as these tend to 
have had more or less constant SFRs during the past 100 Myr, with departures 
that are no more than $3\times$ that from historic means, and so it is then expected 
that galaxies of this type will harbor intrinsically luminous LPVs and massive stars. 
Basic observational properties of these galaxies are summarized in Table 1. 
To the extent that M$_B$ tracks mass then the stellar masses 
of these galaxies agree to within a factor of $2 - 3$. 

\begin{table*}
\begin{tabular}{cccccc}
\tableline\tableline
Name & Type\tablenotemark{a} & B$_T$\tablenotemark{a}\tablenotemark{b} & logD$_{25}$\tablenotemark{a}\tablenotemark{c} & SB$_{D25}$\tablenotemark{a}\tablenotemark{d} & DM$_0$\tablenotemark{e} \\
 & & (mag) & (log arcmin) & (mag arcsec$^{-2}$) & (mag) \\
\tableline
Holmberg II & Im & 11.1 & 1.9 & 15.2 & $27.65 \pm 0.13$ \\
NGC 2366 & IB(s)m & 11.5 & 1.9 & 14.8 & $27.52 \pm 0.28$ \\
IC 2574 & SAB(s)m & 10.8 & 2.1 & 15.3 & $28.02 \pm 0.22$\\
\tableline
\end{tabular}
\caption{Galaxy Properties}
\tablenotetext{a}{From de Vaucouleurs et al. (1991) RC3}
\tablenotetext{b}{Total B brightness.}
\tablenotetext{c}{Isophotal diameter at 25 mag/arcsec$^{2}$ surface brightness in B.}
\tablenotetext{d}{Mean B surface brightness within the D$_{25}$ isophote.}
\tablenotetext{e}Absolute distance modulus from Karachentsev 
et al. (2002), as estimated from the RGB-tip and assuming M$_I = -4.05$ for the 
intrinsic magnitude of the RGB-tip (Da Costa \& Armandroff 1990). The uncertainties 
in these entries are those in the RGB-tip brightness measurements, and do not include 
uncertainties in the reddening and calibration of the RGB-tip brightness.
\end{table*}

	The absolute integrated brightness of Ho II is similar to that of the SMC, 
and HII regions in Ho II have oxygen abundances that are similar to those in 
other dwarf irregular galaxies (Richer \& McCall 1995). 
Gas accounts for 50\% of the total galaxy mass (Puche et al. 1992), and so 
it is not surprising that there is a large population of young blue stars 
that are indicative of very recent large-scale star formation (Hoessel \& Danielson 
1984). Older stars in Ho II are more centrally concentrated than younger stars 
(Bernard et al. 2012), and this is consistent with the tendency for dwarf 
irregular galaxies to have recent star forming activity that is 
more diffusely distributed than at earlier epochs (e.g. Cignoni et al. 2019).

	As in Ho II, there is clear evidence of recent active star formation in NGC 
2366, and a prominent signature of this is the supergiant HII region NGC 2363. NGC 2363 
is in many respects similar to 30 Dor (Kennicutt 1984), and harbors a large population 
of young massive objects, including luminous blue variables (LBVs) and Wolf-Rayet stars 
(e.g. Drissen et al. 2000). Other recent and on-going star formation sites are seen
throughout the galaxy (e.g. Aparacio et al. 1995). McQuinn et al. 
(2010) find evidence for a lull in the SFR 2 -- 3 Gyr ago, although the SFR 
has increased markedly in the past 1 -- 2 Gyr. Thuan \& 
Izotov (2005) identify a population of very bright AGB stars, indicating 
elevated levels of star formation during the past few hundred Myr. Searches for 
red variables in NGC 2366 might then find a number of very luminous 
LPVs. The O abundances of HII regions in NGC 2366 are SMC-like (Talent 1980).

	Based on the B$_T$ and distance modulus entries in Table 1, IC 2574 is the most 
massive of the galaxies. The O abundance deduced from the spectra of HII 
regions is consistent with its M$_B$ (Miller \& Hodge 1996).
Mondal et al. (2019) find evidence of on-going star formation throughout the galaxy. 
Pasquali et al. (2008) examine the photometric properties of IC 2574, and find 
that there is a dominant 100 Myr population within the central 4 kpc, with a 10 Myr 
population at galactocentric radii between 4 and 8 kpc. The diffuse nature of recent star 
formation in IC 2574 is thus qualitatively consistent with the spatial 
distribution of young and old stars in Ho II and other dwarf irregular galaxies. 
Dalcanton et al. (2012) examine the SFH of IC 2574, and 
find that the SFR has increased during the past 20 Myr.

	All three galaxies show evidence of an active interstellar medium (ISM). 
The ISM of Ho II is riddled with cavities that are in excess of a 
kpc across (Puche et al. 1992). Weisz et al. (2009) argue that these may not be 
the result of single concentrated large-scale star formation events, while 
Egorov et al. (2016) suggest that they are large-scale coherent star-forming complexes. 
Cavities similar to those in Ho II are found in the ISM of IC 2574, and star formation 
is occuring along the edges of these structures (Walter \& Brinks 1999).
The largest cavity in IC 2574 is a supergiant shell that has 
massive star-forming regions along its rim that may account for much of 
the present day star formation in that galaxy (Walter et al. 1998; Cannon et al. 2005).
While holes in the ISM are usually attributed to sources internal to a galaxy, 
interactions with high velocity clouds are another possible cause 
(e.g. Mirabel \& Morras 1990). 

	Weisz et al. (2008) investigate the SFHs of all three galaxies by applying 
the same procedure to each system, thereby providing a basis for direct comparison. They 
find that all three galaxies experienced a dip in their mean SFRs during intermediate 
epochs, and that there has been an uptick in the SFRs during the past $\sim 1$ Gyr. 
The SFRs estimated by Weisz et al. (2008) averaged over the past 20 Myr are 
0.08 M$_{\odot}$/year (Ho II), 0.07 M$_{\odot}$/year 
(NGC 2366), and 0.11 M$_{\odot}$/year (IC 2574). 

	HII regions also provide a means of directly comparing the recent 
SFHs of these galaxies. Miller \& Hodge (1994) conduct a census of HII regions in 
all three galaxies, and estimate recent SFRs of 0.05 M$_{\odot}$/year for Ho II, 
0.10 M$_{\odot}$/year for NGC 2366, and 0.08 M$_{\odot}$/year for IC 2574. 
The SFR diagnostics applied by Weisz et al. (2008) and Miller \& Hodge (1994) sample 
slightly different time scales, as HII regions are ionized by stars with ages $\leq 10$ 
Myr, while the Weisz et al. (2008) SFRs are averages over the past 20 Myr. Thus, there 
might be some disagreement between the two estimates.

	There have been previous searches for variable stars in Ho II and NGC 2366, 
and these have identified a number of objects. Hoessel et al. (1998) 
conclude that Ho II is 'extremely rich in variable stars'. They 
identify 28 pulsating variables, of which seven have colors that are consistent with 
those of Cepheids. Tolstoy et al. (1995) identify 13 variable stars in NGC 2366. 
Six are classified as Cepheids and four are probable 
eclipsing binaries. The remaining three were not classified.

	The paper is structured as follows. Information about the images used in this 
study and the processing that was performed on them following the application of 
corrections for instrumental and optical artifacts are discussed in Section 2. 
The procedures used to extract stellar brightnesses are the subject of 
Section 3. Section 4 deals with the identification and characterization of the variables, 
while an examination of their spatial distribution with respect to 
the underlying body of stars follows in Section 5. 
The PLRs of the LPVs are examined in Section 6, and these provide insights into 
the distances of the host galaxies and the pulsational properties of the stars. 
Conclusions and a discussion of the results can be found in Section 7.

\section{ARCHIVAL DATA}

	This study uses images recorded for the SPIRITS survey 
(Kasliwal et al. 2017), the goal of which is to identify photometric 
transients in {\it Spitzer} IRAC (Fazio et al. 2004) images of nearby galaxies. 
Survey galaxies were typically observed for 10 epochs over a 
$\sim 2$ year time span, although additional exposures 
with longer time baselines were recorded for some galaxies. 
Images for each epoch have a total exposure time of 93.6 sec, and these 
are the deepest IRAC observations of the galaxies considered in the present study.

	Processed [3.6] and [4.5] images of the galaxies 
were downloaded from the IPAC {\it Spitzer} Heritage Archive 
\footnote[1]{https://sha.ipac.caltech.edu/applications/Spitzer/SHA/}. 
Two sets of images are available for Ho II: one is centered on the 
main body of the galaxy, while the other is centered on 
Ho II X-1 (Zezas et al. 1999), which is located to one side of the 
galaxy. Only the images in the first group are considered here. 
The AORs of the images used in this study are listed in Table 2.

\begin{table*}
\begin{center}
\begin{tabular}{ccc}
\tableline\tableline
Ho II & NGC 2366 & IC 2574 \\
\tableline
50639104 & 50508800 & 50580992 \\ 
50639872 & 50509056 & 50581504 \\ 
50640384 & 50509312 & 50582016 \\ 
52733184 & 52721600 & 52720640 \\ 
52733440 & 52729856 & 52720896 \\ 
52733696 & 52730112 & 52721152 \\ 
52733952 & 52730368 & 52721408 \\ 
52734208 & 52730624 & 52721664 \\ 
52734464 & 52730880 & 52721920 \\ 
52734720 & 52731136 & 52722176 \\ 
\tableline
\end{tabular}
\caption{Astronomical observation requests (AORs) for images used in this study}
\end{center}
\end{table*}

	The observations of each galaxy were recorded with a range of 
satellite position angles. The downloaded images for each galaxy were 
rotated to a common orientation, aligned, and then trimmed to the area of 
common sky coverage. While the last step curtails the areal coverage, 
it ensures that a complete set of epochs is available for each 
star, thereby producing a homogeneous dataset for period determination and 
characterization. A background light level was then measured 
and subtracted from each aligned image. These background-subtracted 
images for each galaxy$+$filter combination 
form the final datasets that are examined for stellar variability.

	Reference [3.6] and [4.5] images for each galaxy were constructed by finding 
the median light level at each shifted pixel location in the background-subtracted 
images. The brightnesses of variable stars in these median images provide 
'typical' values over the observing timespan, and so are a reference for assessing 
variability and obtaining representative magnitudes and colors for gauging the 
evolutionary state of each star. The median [3.6] image of each galaxy is shown in 
Figure 1.

\begin{figure}
\figurenum{1}
\plotone{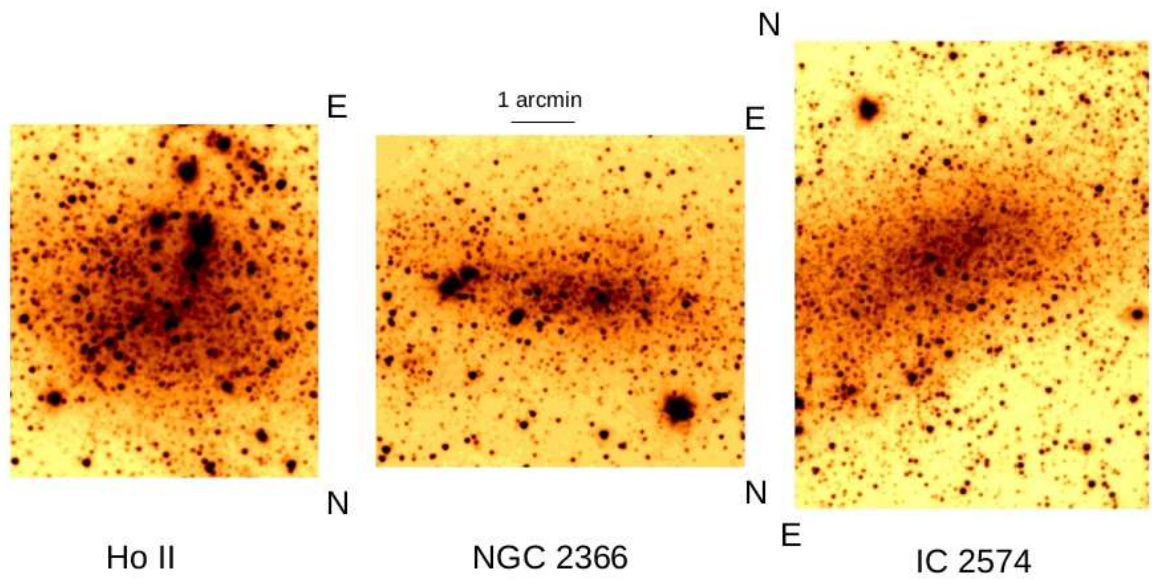}
\caption{Median [3.6] images, obtained by aligning the ten 
background-subtracted exposures of each galaxy and then finding the median signal at each 
aligned pixel location. The results were then trimmed to the area of common sky coverage 
as defined by the ten exposures recorded for each galaxy. The approximate locations of 
North and East are marked.}
\end{figure}

\section{PHOTOMETRIC MEASUREMENTS}

	Angular resolution elements in the [3.6] and [4.5] images of these 
galaxies subtend a projected area of $\sim 1000$ pc$^2$, and so source blending 
is a concern when measuring stellar brightnesses. These concerns 
are mitigated to some degree by the intrinsic IR brightnesses of LPVs, 
which are such that they stand out from the surrounding background of 
unresolved stars. The effect of this background on the detection of variables can be 
suppressed by differencing images taken with the same exposure time in the same 
filter with the same angular resolution. This is not a new concept: Tomaney \& Crotts 
(1996) discuss the detection of variable stars in crowded M31 fields, and demonstrate 
that reliable light variations can be detected differentially after suitable adjustments 
are made for variations in the point spread function (PSF) between exposures and within 
the science field. Contrast arguments aside, there are also areas in crowded environments 
where stochastic effects may produce angular resolution elements 
where the stellar density is comparatively low, with the result that the light 
in a given resolution element may originate predominantly from a single source, 
even in areas with extremely high projected mass densities (Davidge et al. 2010).

	Kasliwal et al. (2017) point out that the PSF is poorly sampled in 
raw IRAC images, and this could result in false detections. In recognition of this, 
photometry for the SPIRITS database was conducted by making aperture measurements 
on differenced images. While the prognosis for obtaining 
reliable changes in light levels from aperture measurements 
is promising, there is a tendency for at least some of the variables to be located 
in locally dense environments. This complicates attempts to obtain the reliable 
{\it absolute} photometric measurements that are required to place these objects on CMDs. 
In any event, photometry of non-variable sources in a galaxy is required to 
understand the stellar content of the host system and detect objects of 
special interest, such as the very massive stars that are also a target 
of this study. Finally, if a star in close proximity to a variable is itself variable 
then this may compromise the photometry of both objects made from differenced images. 

	PSF sampling issues aside, Davidge (2014) shows that the 
application of PSF-fitting to IRAC images of crowded fields 
results in photometric measurements that are deeper 
and have smaller uncertainties than those obtained from aperture measurements. Given the 
crowded nature of some parts of the galaxy images, the photometric measurements 
used here were made with the PSF-fitting program ALLSTAR (Stetson 
1994). Source catalogues, preliminary magnitudes, 
and PSFs were constructed from the median [3.6] and [4.5] images 
using routines in DAOPHOT (Stetson 1987). The PSFs were 
constructed from bright, unsaturated, isolated stars, and stellar brightnesses were 
found by fitting the PSFs to the area within the FWHM (full width at half maximum) in 
stellar light profiles. The source catalogues obtained from the [3.6] images 
were adopted to perform photometry on the [4.5] images. 
The source catalogues, initial magnitudes, and PSFs constructed from 
the median images were then used to perform photometry on the exposures 
recorded at different epochs.

	The unresolved body of stars in these galaxies forms a 
non-uniform background that complicates efforts to determine 
reliable local background light levels. This is of greatest concern for objects 
that are near the photometric faint limit, where background light is a significant 
fraction of the observed light within the PSF. Faint stars are thus most susceptible 
to uncertainties in the local background. The presence of this background is not an issue 
for photometry made from differenced images.

	A template of the unresolved galaxy background was constructed and 
removed from the images to account for non-uniform local light levels. 
Preliminary magnitudes were first measured from images with 
the background light in place. All photometered 
stars were then subtracted from the images, and the results were smoothed 
with a $20 \times 20$ arcsec median filter to suppress residuals from the 
subtraction process. The smoothing kernel size was selected as a compromise 
between the need to suppress subtraction artifacts while still retaining angular 
resolution in the template. The resulting smoothed images track the 
unresolved background light, and were subtracted from the initial images. A final 
set of magnitudes was then obtained from the background-subtracted images. The removal 
of the background light improved the photometry in two ways. First, the number of stars 
that were retained by ALLSTAR was increased by $\sim 10\%$. Second, the 
dispersion in measurements near the faint end of the CMDs was also reduced.

	Examples of the photometry obtained using the procedures described above 
as well as the suppression of systematic effects when measuring light variations  
are shown in Figure 2. The left hand panel of Figure 2 shows the $([3.6],[3.6]- [4.5])$ 
CMD of NGC 2366. The [3.6]--[4.5] color broadens when [3.6]$ > 16$, marking 
the onset of large numbers of evolved stars in NGC 2366 that have hot circumstellar dust 
envelopes. In Section 4.1 it is shown that this is also the approximate magnitude where 
the number of sources detected in the median NGC 2366 image exceeds that expected from 
foreground and background sources; still some objects with [3.6] $< 16$ may 
belong to NGC 2366. 

\begin{figure}
\figurenum{2}
\plotone{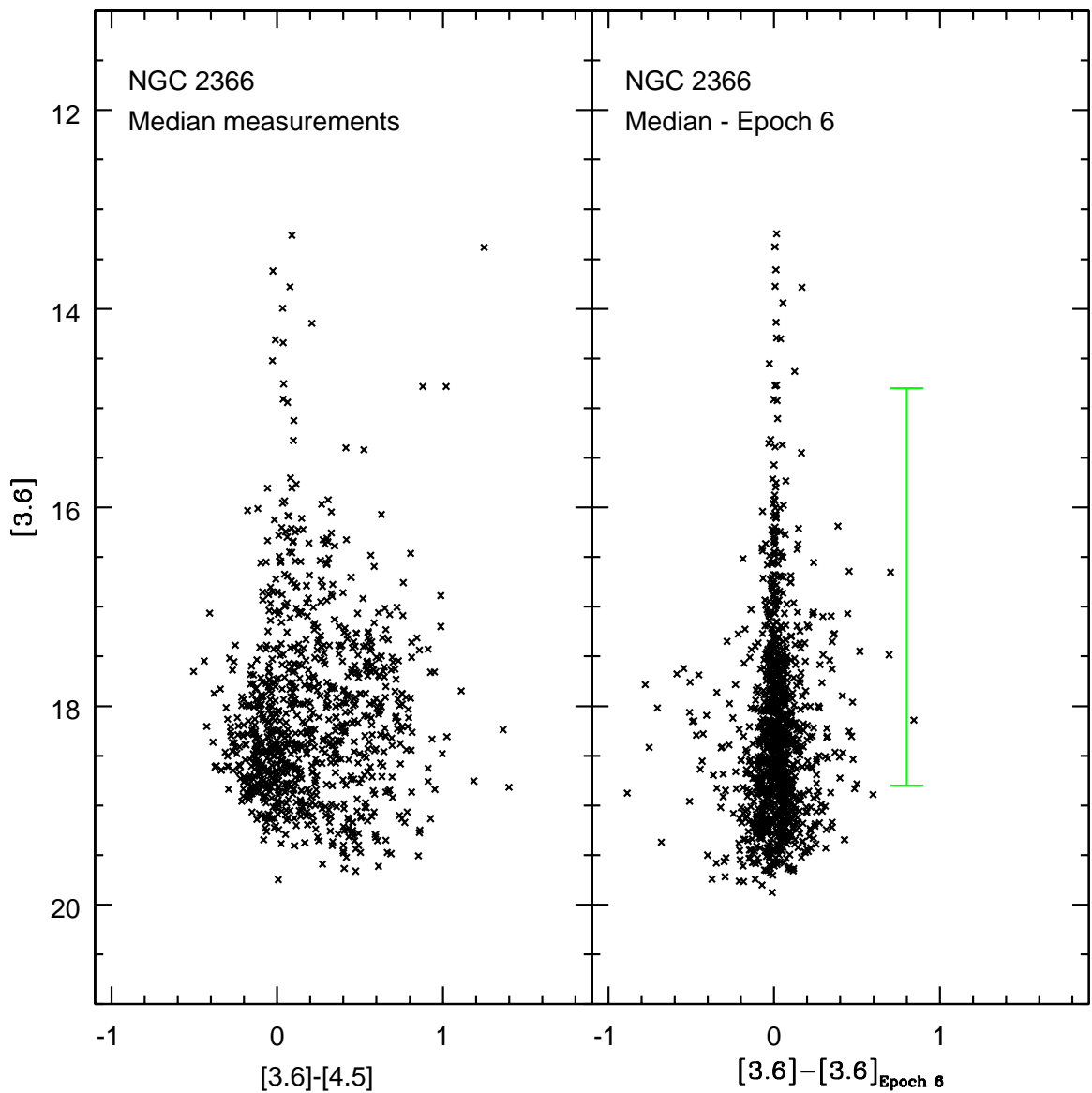}
\caption{Example of the absolute and differential photometric measurements 
obtained in this study. Left hand panel: $([3.6], [3.6]-[4.5])$ CMD of NGC 2366, based on 
photometric measurements obtained from the median images. The CMD broadens when [3.6] $> 
16.5$, which is the magnitude range where stars in NGC 2366 occur in significant 
numbers. Right hand panel: The difference between [3.6] magnitudes 
obtained from the median and Epoch 6 (see text) images. The range of 
median magnitudes that were searched for variable stars is indicated 
with the green line. The objects that fall to the right and 
left of the dominant plume are possible variable stars. 
The narrow dispersion along the x axis in the right hand panel demonstrates that 
differential photometric measurements approaching the few percent level are 
recovered from these data.}
\end{figure}

	The suppression of systematic effects is demonstrated in the right hand panel 
of Figure 2, which shows the difference in [3.6] magnitudes between measurements 
made from the AOR 52730112 (hereafter 'Epoch 6') image and median-combined images 
for each star. In contrast to the CMD in the left hand panel, the differences between the 
measurements made from the Epoch 6 and median-combined datasets form a narrow sequence 
that widens progressively towards fainter magnitudes, with a width that amounts to 
only a few hundredths of a magnitude over a wide range of [3.6] magnitudes, even 
near the faint limit. The compact width of this sequence is noteworthy given the 
range of stellar densities in the NGC 2366 image (Figure 1). The spray of points that 
cluster about the dominant plume in the right hand panel of Figure 2 are potential 
variable stars. The tight sequence in the right hand panel of Figure 2 demonstrates the 
feasibility of identifying modest changes in light levels, even near the faint limit of 
the data. Similar results are obtained when measurements from other epochs are examined. 

	The photometric measurements obtained here are 
deeper and have a lower random uncertainty than those previously 
published. For example, Williams \& Bonanos (2016) present 
IRAC photometry of 505 sources in NGC 2366 over a 28.8 arcmin$^2$ area. 
The $([4.5], [3.6]-[4.5])$ CMD of NGC 2366 obtained from the Williams \& Bonanos 
(2016) photometry is compared with the CMD obtained from our measurements in Figure 3. 
While the two CMDs have similarities when [4.5] $< 15$, at fainter 
magnitudes they are very different, in the sense that the CMD constructed from the 
measurements made here is better defined and contains many more stars. In particular, we 
recover in excess of 6500 sources in NGC 2366 over an area of 22.6 arcmin$^2$. 

\begin{figure}
\figurenum{3}
\plotone{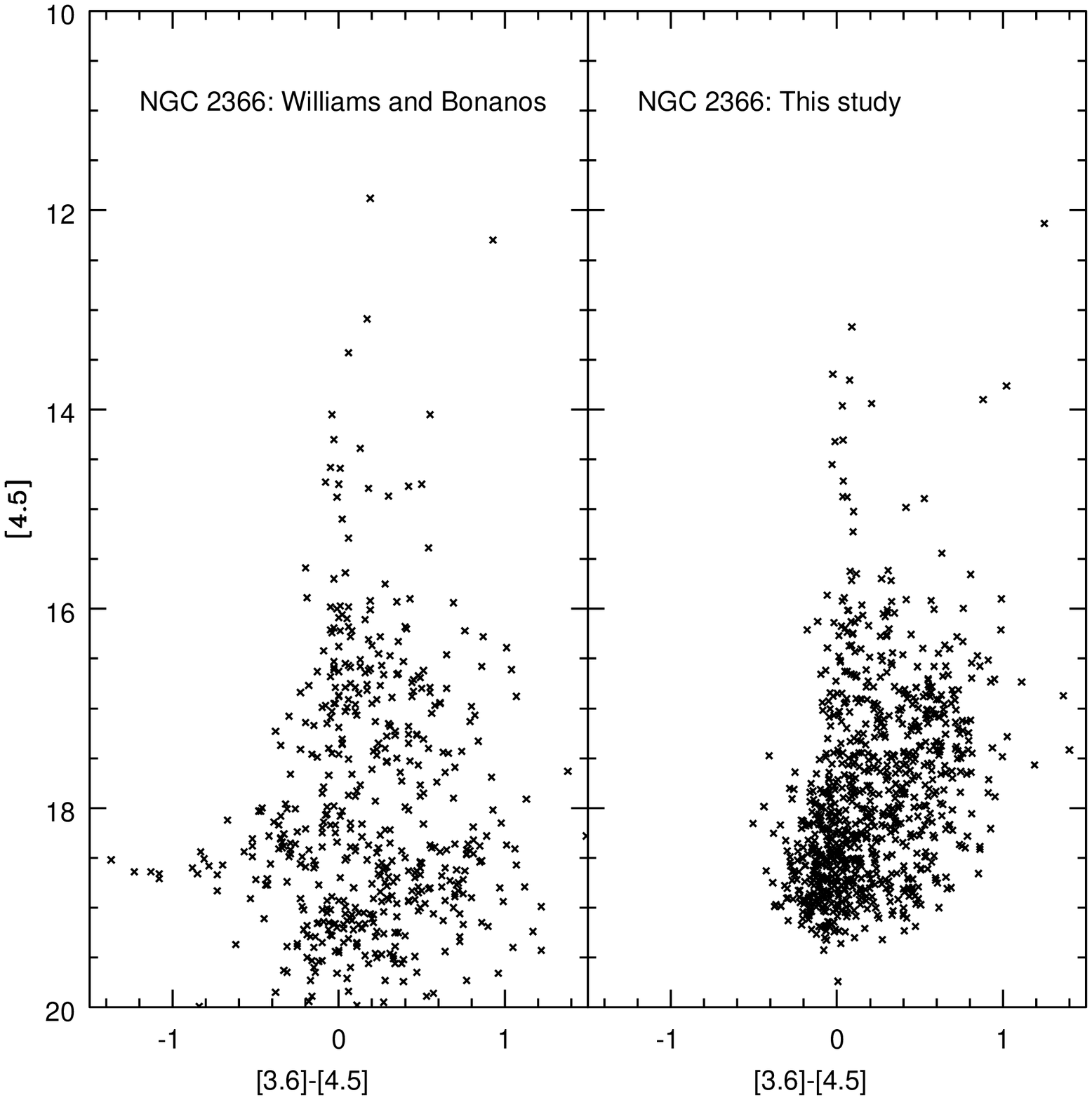}
\caption{$([4.5], [3.6]-[4.5])$ CMDs of stars in NGC 2366 obtained 
from the median SPIRITS images and from the IRAC images photometered by 
Williams \& Bonanos (2016). While there are similarities between the two CMDs when 
[3.6] $< 15$, at fainter magnitudes the larger number of points and the smaller 
scatter in colors in the right hand panel indicate that the measurements made in the 
current study have smaller uncertainties at magnitudes where stars in NGC 
2366 are present than those in the Williams \& Bonanos (2016) measurements.} 
\end{figure}

\section{DETECTION AND CHARACTERIZATION OF VARIABLEs}

\subsection{Identification}

	The search for variable stars in the current study was conducted within a 
[3.6] magnitude interval that covers the brightest stars in each galaxy, 
as well as stars at fainter magnitudes where it was thought that complete light curves 
could be obtained without interference from the photometric faint limit. 
Variability is assessed by comparing differenced photometric measurements 
obtained from PSF-fitting (see previous section). The use of a single source 
catalogue to measure stellar brightnesses in each set of images, coupled with the 
inherent optical stability of a space-based telescope, greatly helps to 
suppress systematic effects that might otherwise crop up due to crowding when measuring 
image-to-image changes in photometric measurements. 

	Two different techniques are used to identify variables. Davidge (2021) 
surveyed the southern disk of NGC 247 at visible wavelengths 
to search for variable stars, and the procedures discussed in that paper 
are applied here. The primary criterion to identify variables employed by Davidge (2021) 
is the standard deviation ($\sigma$) in differenced magnitude measurements. A star 
is deemed to be variable if the standard deviation in the measurements made at different 
epochs differs from that defined by non-variables with the same 
magnitude in the median image at the $5\sigma$ or higher level. 
While strict, the $5\sigma$ criterion ensures that fluctuations due to noise are 
not flagged as variables. The dispersion for non-variables was estimated from the 
[3.6]$_{Epoch X}$--[3.6]$_{median}$ measurements, where [3.6]$_{Epoch X}$ 
is the brightness measured at epoch 'X', and [3.6]$_{median}$ is the brightness 
in the median image. An iterative rejection filter was applied to remove 
the contribution made by variable stars when computing the non-variable dispersions.

	Another requirement was recovery in all ten images 
recorded for each galaxy. This ensured light curves that are as 
complete as possible, which is beneficial for estimating periods in a 
consistent manner and classifying the nature of the variability given the spotty 
phase coverage. Eruptive variables with episodes that last only a few days and that 
have quiescent states below the photometric faint limit are then not flagged as 
variables, as they are not detected in the full suite of images.

	Sokolovsky et al. (2017) discuss and evaluate 
various techniques that can be applied to detect photometric variability. 
A robust technique identified in their study uses the median absolute 
deviation (MAD) of photometric measurements, and a detection scheme that uses this 
statistic is the second detection technique that is applied to the IRAC images. 
As the MAD is a non-parametric statistic, it is robust to outliers, and so defines 
a detection criterion that is sensitive to variables 
with cyclical light curves or that show frequent non-periodic excursions. It is 
distinct from the $\sigma$ criterion described previously, which relies on a normal 
distribution of data points and is susceptible to outliers.

	When applying the MAD detection technique to the [3.6] images of the 
three galaxies, a threshold that is $5\times$ the MAD 
measured for non-variable objects was applied for consistency with the $\sigma$ detection 
threshold. With this threshold, the MAD technique recovers numbers of variables 
in all three galaxies that are comparable to those found with the $\sigma$ technique, 
suggesting that the $\sigma$ and MAD thresholds are consistent. 
As with the $\sigma$ technique, a variable star had to be recovered in all ten images 
of a galaxy.

	The $\sigma$ and MAD-based techniques detect many variables in common. 
However, 15 variables were found with the MAD criterion that were not found 
with the $\sigma$ criterion. The variables found 
with MAD are unremarkable when compared with the 
variables found with the $\sigma$ criterion. All are LPVs with well-defined 
light curves that have periods and amplitudes that are consistent with those 
found using the $\sigma$ criterion. The variable star sample discussed in the 
remainder of the paper is the union of the objects identified with the 
$\sigma$ and MAD-based techniques.

	An examination of the variables that are not common to both 
techniques reveal that these objects tend to be borderline detections. 
One reason for this is that the magnitude measured from the median image, 
which is adopted as the reference magnitude for the $\sigma$ detection scheme, is not 
the same as the median magnitude from the various epochs that is used for the reference 
magnitude by the MAD technique. While these two reference magnitudes typically differ 
by only a few percent, some variables fall slightly above the faint limit 
of the search with one technique but not the other. Similarly, while the standard 
deviation and MAD tend to be consistent (the $\sigma$ is expected to be $1.483\times$ 
the MAD for a normal distribution) there are subtle differences, causing some stars 
to fall slightly above the 5 deviation detection threshold for one technique but 
not the other. The differences between the samples of variables detected by the two techniques 
are thus attributed to subtle differences in the adopted thresholds and reference 
magnitudes, and not to systematic differences in the characteristics of the 
light curves.

\subsection{Light Curves, and Periods}

	Light curves of the variables detected in the IRAC images are shown in Figures 4, 
5, and 6. Ho II and NGC 2366 have been surveyed previously for variables stars 
(Tolstoy et al. 1995; Hoessel et al. 1998), and a number of objects were identified 
in both galaxies. To avoid confusion with the naming schemes assigned 
in those surveys, the variables found here have been assigned a 'DV' 
(Davidge Variable) number.

	The majority of the light curves exhibit the rhythmic, large 
amplitude light variations that are characteristic of LPVs. 
Among those objects that are likely LPVs there is a tendency for the brighter 
variables to have longer periods (i.e. a smaller number of cycles 
sampled by these data), as expected given that there is a PLR for LPVs that are pulsating in 
the fundamental mode. There are also some variables, such as DV28 and DV33 
in IC 2574 that are likely not LPVs, and spectroscopic follow-up may find 
that some of these are highly evolved massive stars. 

\begin{figure}
\figurenum{4a}
\plotone{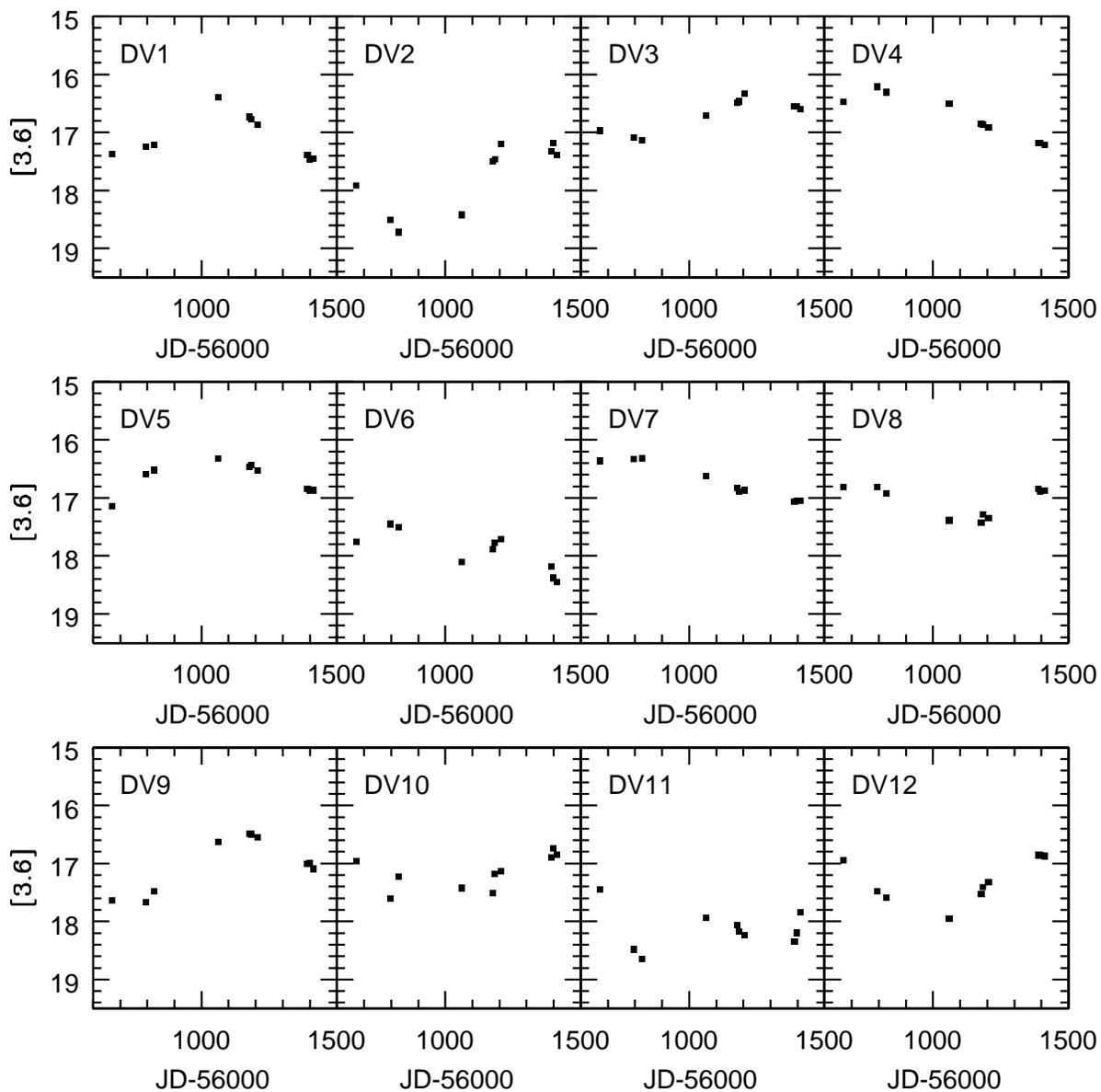}
\caption{Light curves of variable stars in Ho II. Many of the 
light curves show the rhythmic behaviour that is characteristic of LPVs. There is a 
tendency for shorter period variables (i.e. those with multiple variation cycles 
in the light curves) to occur at fainter magnitudes, as expected given the PLR for LPVs 
that are pulsating in the fundamental mode. In fact, while the light curves have spotty phase 
coverage, the periods estimated from these data are sufficiently reliable to obtain a 
reasonably tight PLR for Ho II (Section 6).}
\end{figure}

\begin{figure}
\figurenum{4b}
\plotone{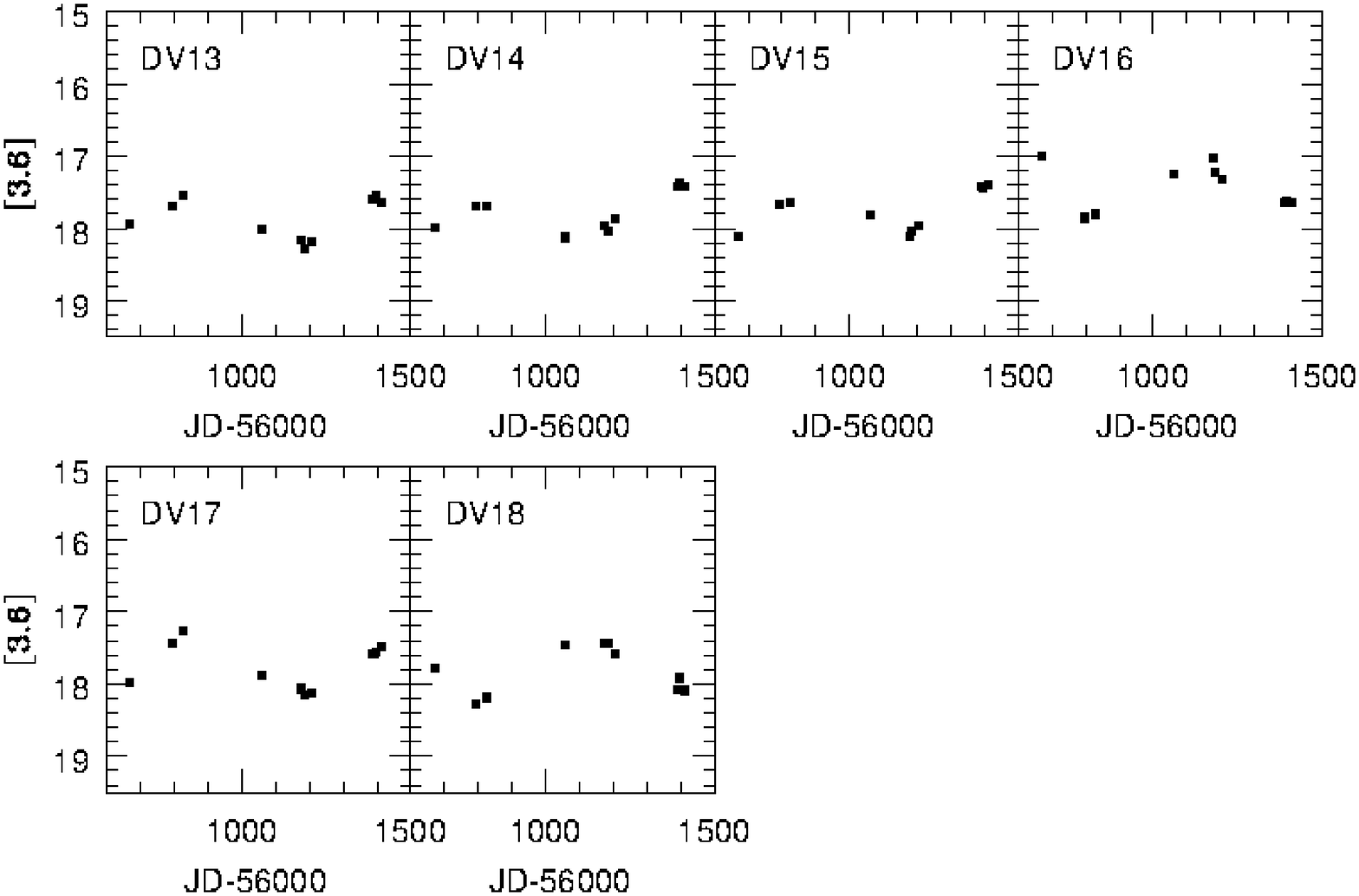}
\caption{Figure 4 (con't).}
\end{figure}

\begin{figure}
\figurenum{5a}
\plotone{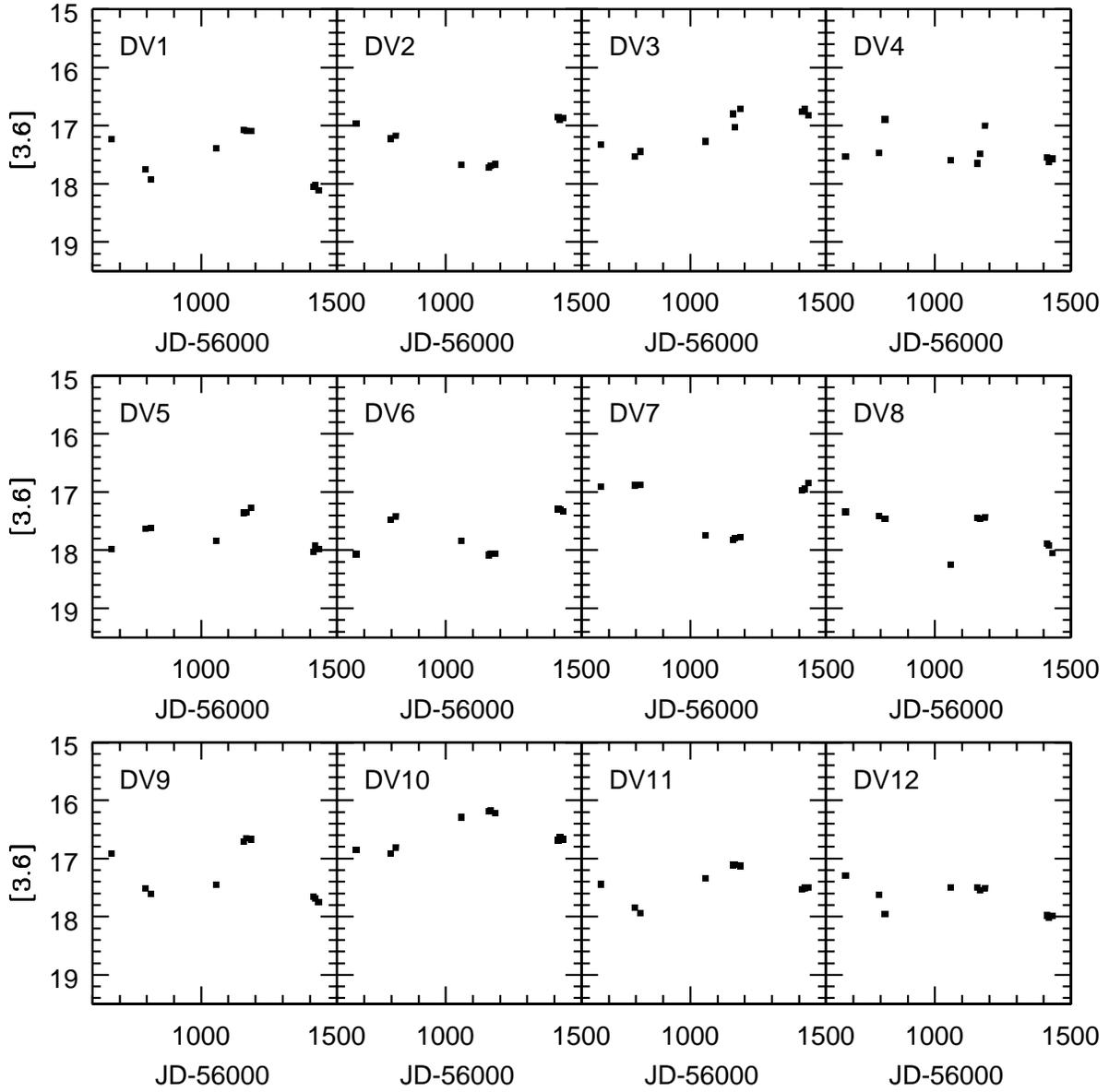}
\caption{Same as Figure 4, but showing light curves of variable stars in NGC 2366.}
\end{figure}

\begin{figure}
\figurenum{5b}
\plotone{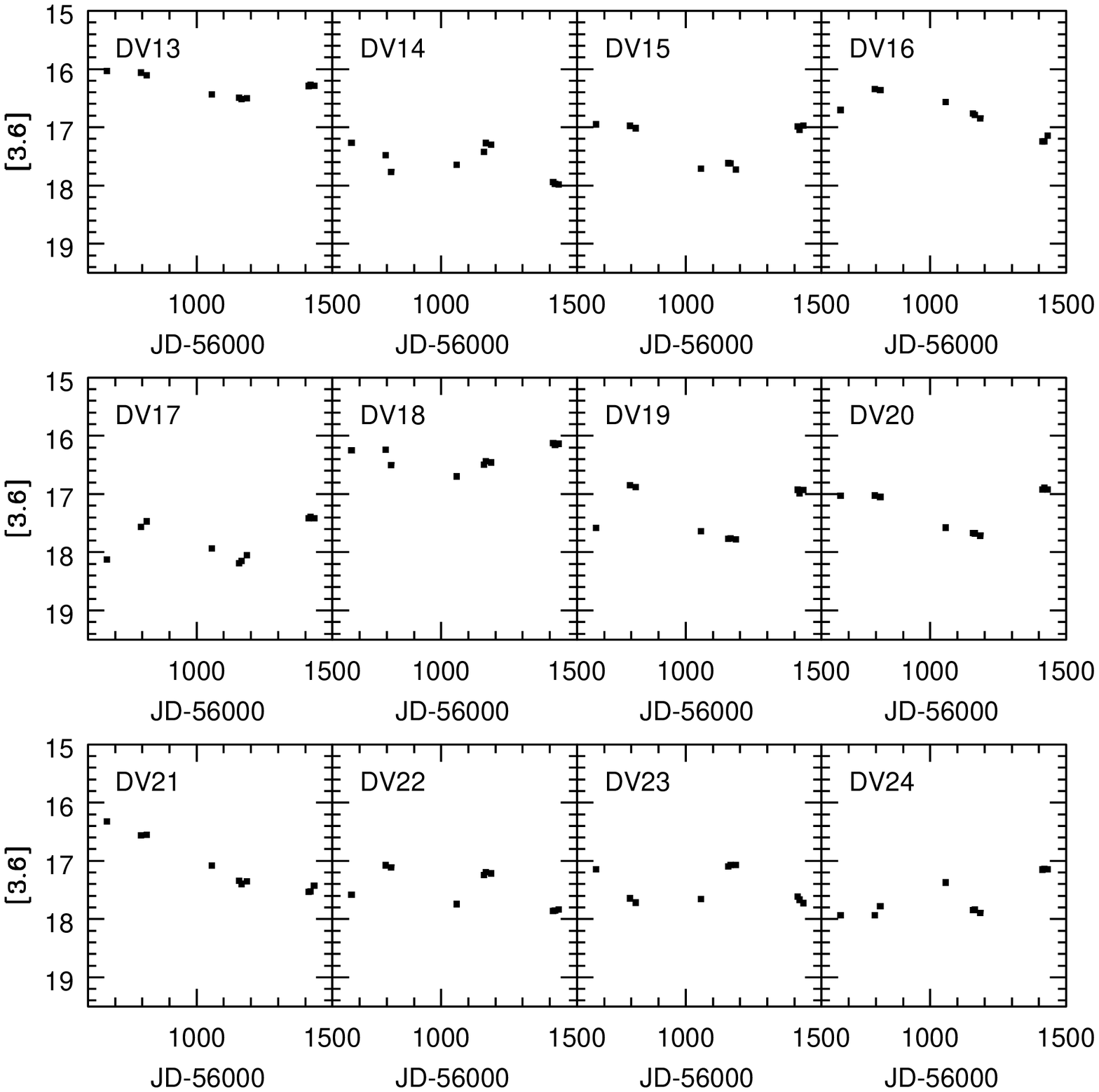}
\caption{Figure 5 (con't).}
\end{figure}

\begin{figure}
\figurenum{5c}
\plotone{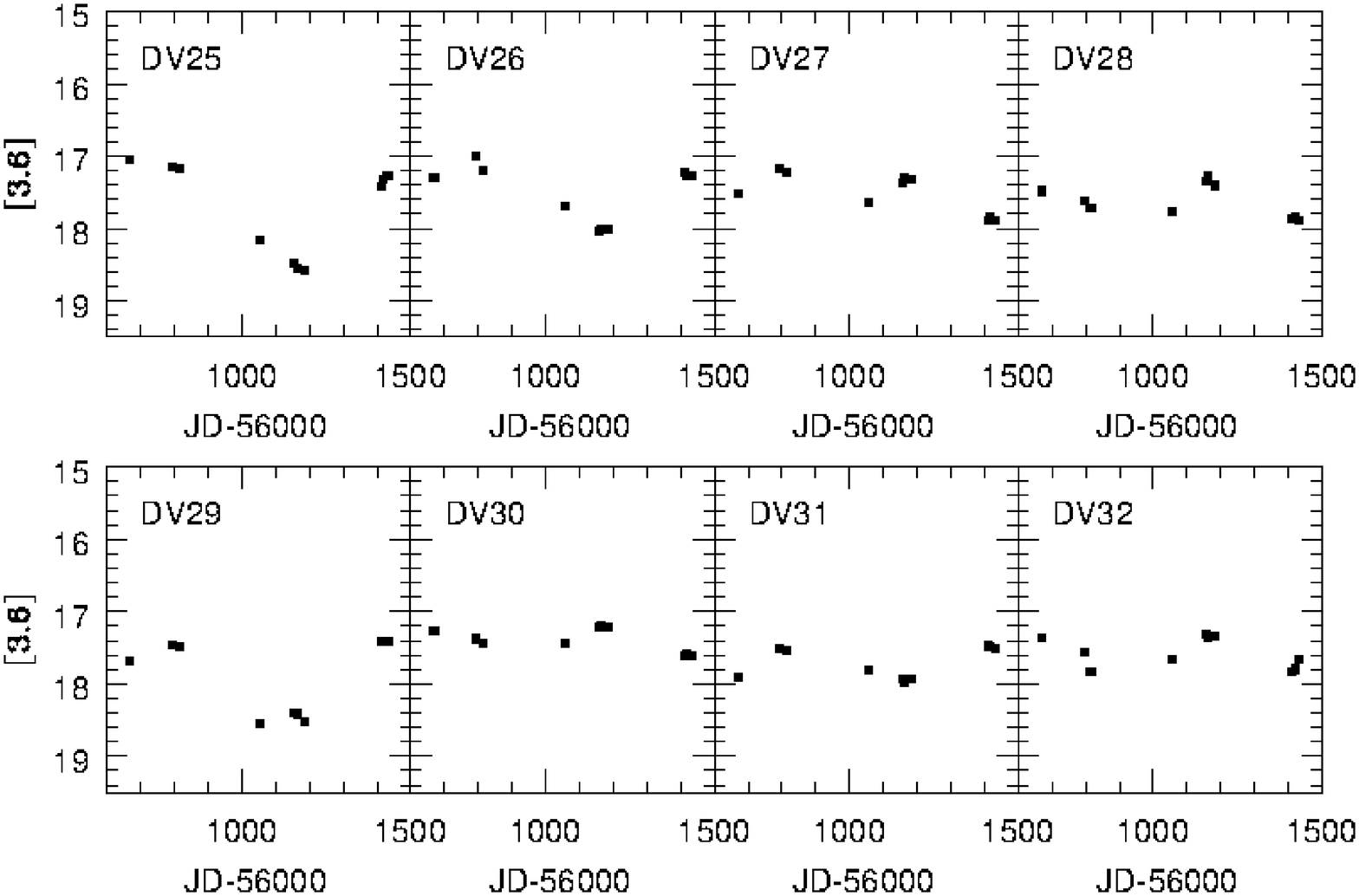}
\caption{Figure 5 (con't).}
\end{figure}

\begin{figure}
\figurenum{6a}
\plotone{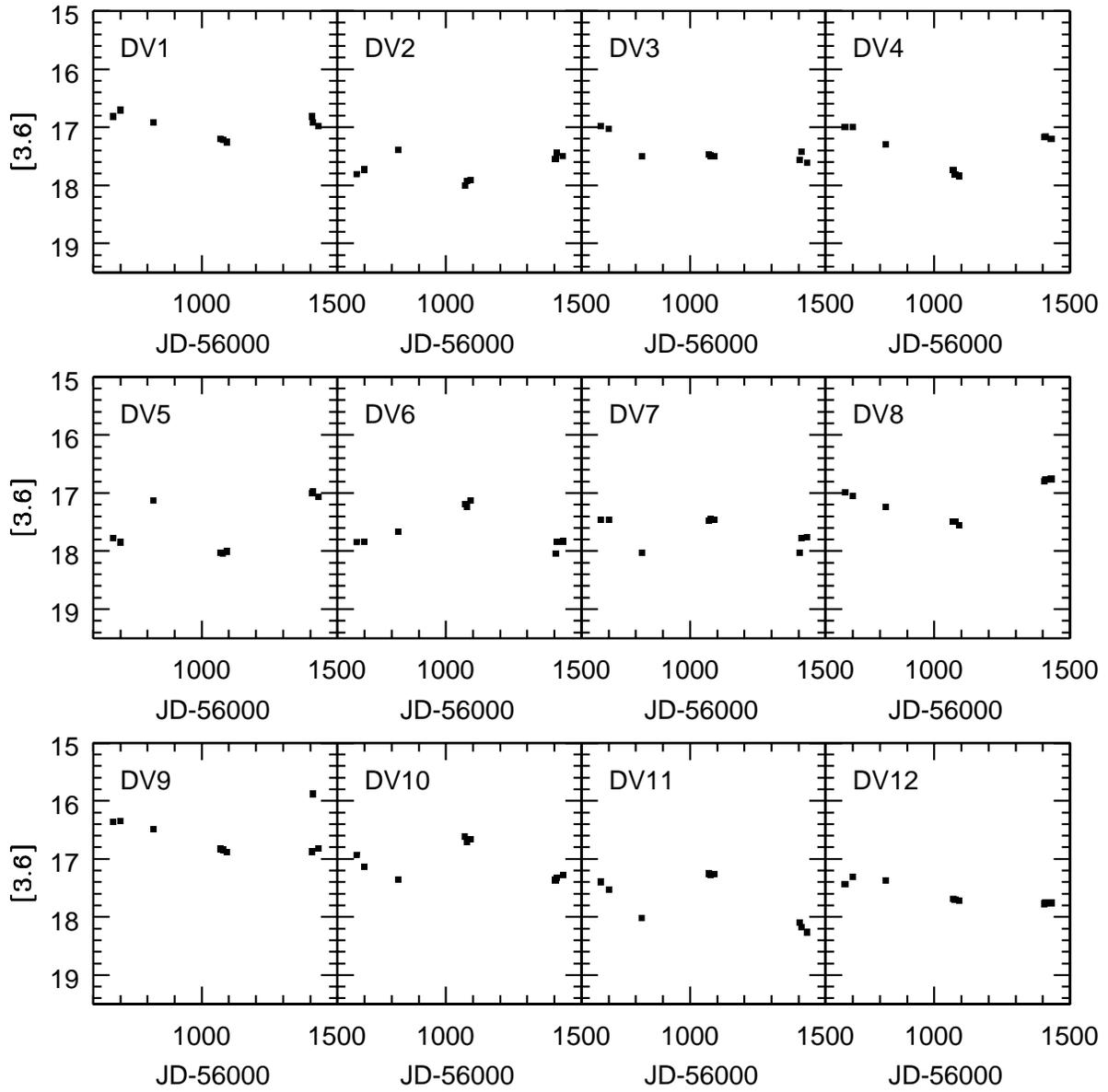}
\caption{Same as Figure 4, but showing light curves of variable stars in IC 2574}
\end{figure}

\begin{figure}
\figurenum{6b}
\plotone{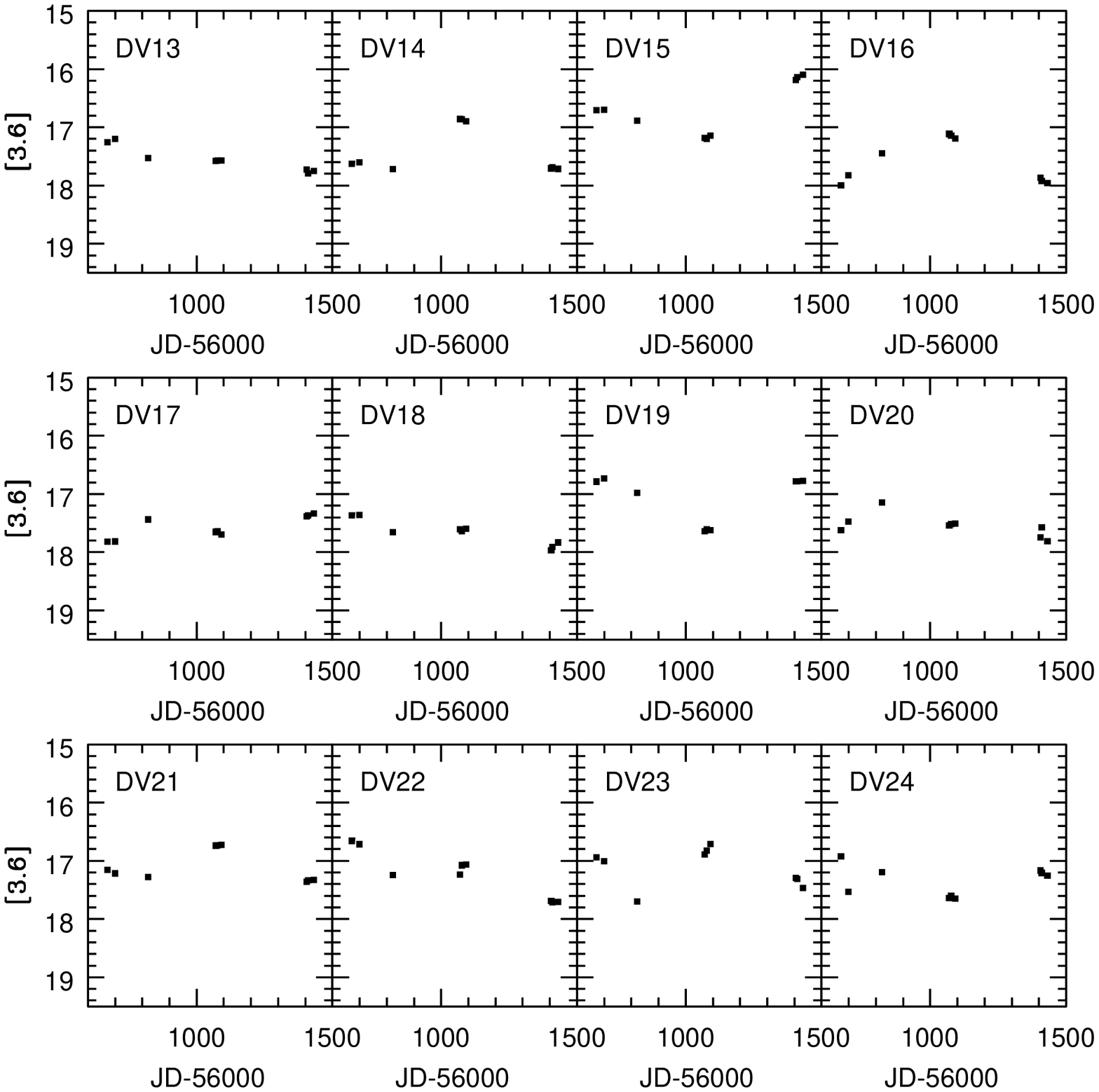}
\caption{Figure 6 (con't)}
\end{figure}

\begin{figure}
\figurenum{6c}
\plotone{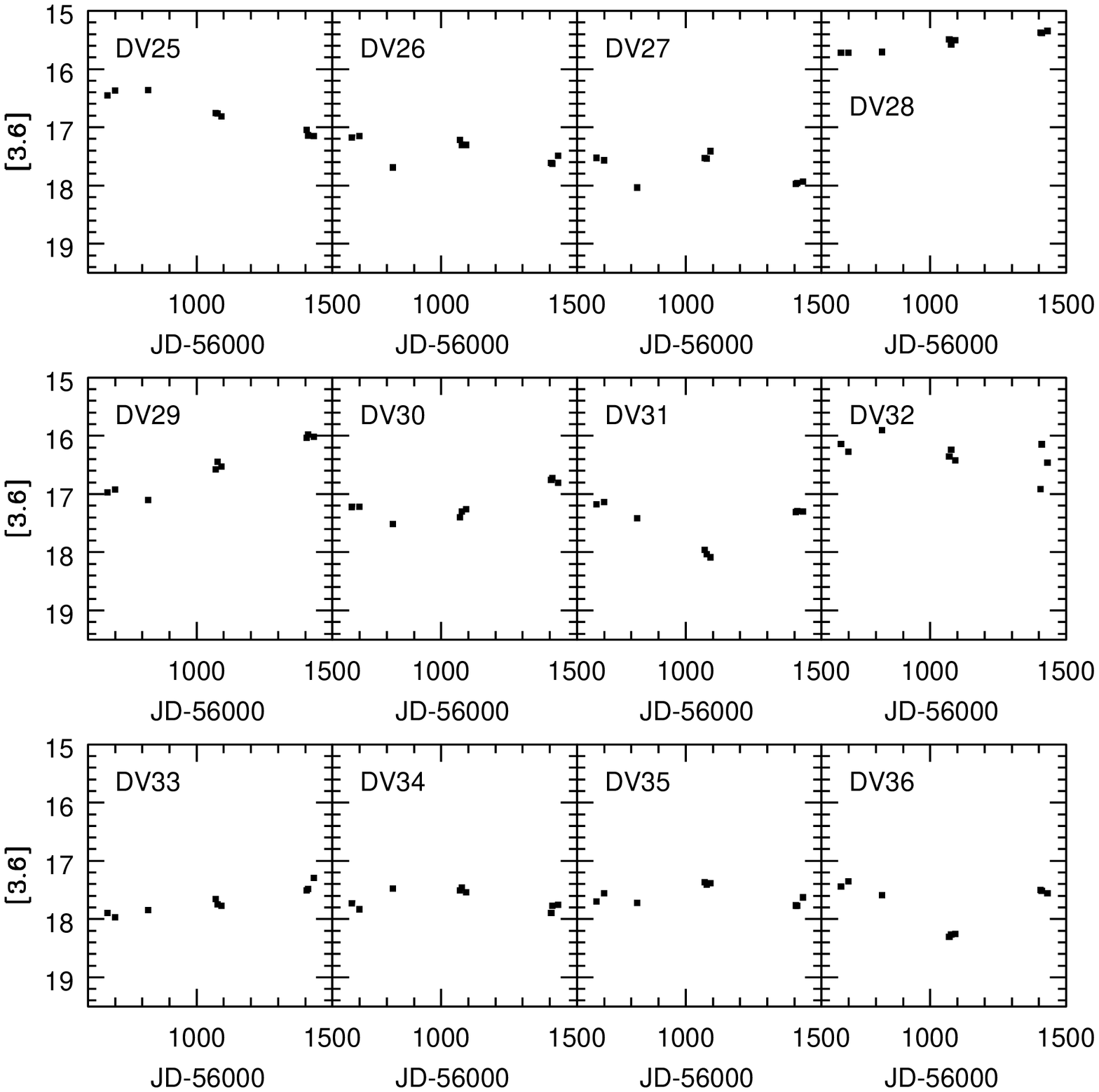}
\caption{Figure 6 (con't)}
\end{figure}

\begin{figure}[!ht]
\figurenum{6d}
\epsscale{0.35}
\plotone{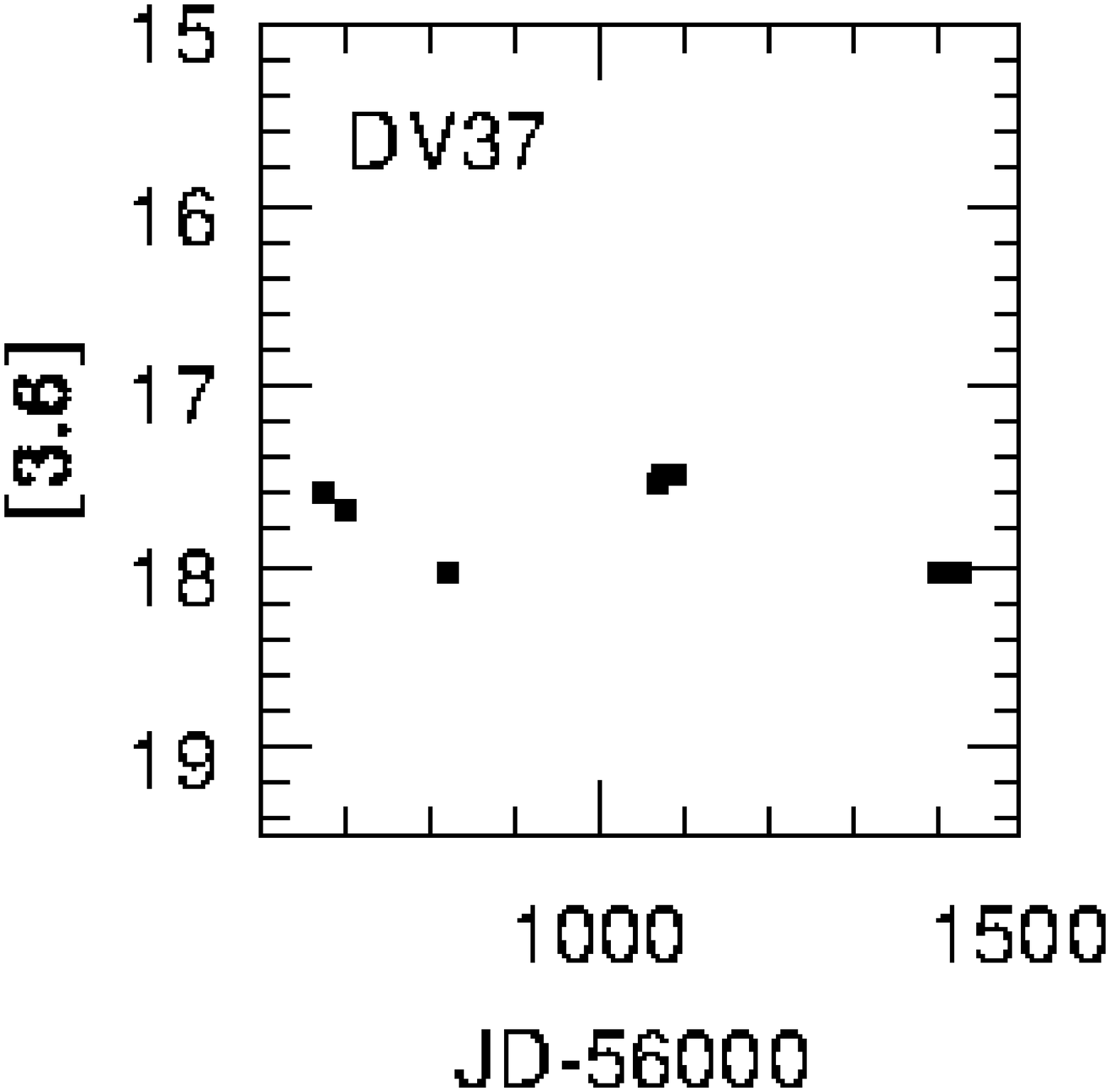}
\caption{Figure 6 (con't)}
\end{figure}

	On-sky co-ordinates, preliminary periods, characteristic magnitudes, 
colors, and amplitude measurements are listed in Tables 3, 4, and 5. 
The photometric measurements are from the median images, while the photometric errors 
given in the table footnotes are those in the photometric calibration. These 
are based on comparisons with the brightnesses and colors of point sources in the 
NASA/IPAC Infrared Science Archive. The amplitude estimates are the difference between 
the brightest and faintest points in each light curve, and so are lower limits to the 
intrinsic photometric amplitude as there are gaps in the light curves. However, Goldman et al. 
(2019) examine the relation between the light curve amplitude of LPVs and [3.6]--[4.5] 
color. When binned according to amplitude, there is a general tendency for 
the mean [3.6]-[4.5] colors in these tables to increase with amplitude and follow the general 
relationship found by Goldman et al. (2019). This suggests that the amplitude measurements 
are reliable proxies of the true intrinsic amplitudes. The 
classification listed in the last column is based on the shape of the light 
curve and location on the CMD; these are discussed at greater length in the 
next section. A question mark indicates that the classification is preliminary 
only. The right ascensions and declinations listed in these tables use 
the World Co-ordinate System information in the image headers. 

\begin{table*}
\begin{center}
\begin{tabular}{cccccccc}
\tableline\tableline
DV\# & RA & Dec & Period & [3.6]\tablenotemark{a} & [3.6]--[4.5]\tablenotemark{a} & Amplitude\tablenotemark{b} & Type \\
 & (J2000) & (J2000) & (days) & (mag) & (mag) & (mag) & \\
\tableline
1 & 08:19:02.9 & 70:44:51 & -- & 17.11 & 0.57 & 1.1 & LPV? \\
2 & 08:18:47.0 & 70:42:33 & 800\tablenotemark{c} & 17.52 & 1.06 & 1.6 & LPV \\
3 & 08:19:03.2 & 70:43:25 & 900\tablenotemark{c} & 16.60 & 0.39 & 0.8 & LPV \\
4 & 08:19:15.0 & 70:44:05 & -- & 16.83 & 0.33 & 1.0 & LPV \\
5 & 08:18:58.9 & 70:42:20 & -- & 16.60 & 0.34 & 0.8 & LPV \\
6 & 08:19:02.1 & 70:42:20 & 500 & 17.83 & 0.78 & 1.1 & LPV \\
7 & 08:19:01.5 & 70:42:07 & -- & 16.78 & 0.28 & 0.6 & LPV \\
8 & 08:18:58.0 & 70:41:36 & 700 & 16.94 & 0.51 & 0.5 & LPV \\
9 & 08:19:09.6 & 70:42:13 & -- & 16.98 & 0.75 & 1.2 & LPV? \\
10 & 08:19:18.0 & 70:43:00 & 800\tablenotemark{c} & 17.04 & 0.23 & 0.8 & RSG? \\
11 & 08:19:09.2 & 70:42:00 & 450 & 18.17 & 0.92 & 1.2 & LPV \\
12 & 08:19:22.4 & 70:43:22 & 700 & 17.33 & 0.86 & 1.1 & LPV \\
13 & 08:19:10.0 & 70:41:14 & 550 & 17.80 & 0.66 & 0.8 & LPV \\
14 & 08:18:47.7 & 70:44:08 & 500 & 17.74 & 0.24 & 0.8 & LPV \\
15 & 08:19:10.9 & 70:44:03 & 550 & 17.72 & 0.54 & 0.9 & LPV \\
16 & 08:19:29.9 & 70:43:32 & 450 & 17.46 & 0.20 & 0.9 & LPV \\
17 & 08:19:23.1 & 70:44:04 & 500 & 17.65 & 0.58 & 0.9 & LPV \\
18 & 08:19:21.1 & 70:44:22 & 650 & 17.72 & 0.66 & 0.7 & LPV \\
\tableline
\end{tabular}
\end{center}
\caption{HoII Variable Stars}
\tablenotetext{a}{Measured from median images. The estimated uncertainties 
are $\pm 0.06$ magnitude for [3.6], and $\pm 0.03$ magnitude for [3.6]--[4.5].}
\tablenotetext{b}{Difference between the brightest and faintest points.}
\tablenotetext{c}{Complete cycle not observed.}
\end{table*}

\begin{table*}
\begin{center}
\begin{tabular}{cccccccc}
\tableline\tableline
DV\# & RA & Dec & Period & [3.6]\tablenotemark{a} & [3.6]--[4.5]\tablenotemark{a} & Amplitude\tablenotemark{b} & Type \\
 & (J2000) & (J2000) & (days) & (mag) & (mag) & (mag) & \\
\tableline
1 & 07:28:42.9 & 69:14:38 & 550 & 17.44 & 0.86 & 1.0 & LPV \\
2 & 07:28:24.8 & 69:11:53 & 750 & 17.24 & 0.22 & 0.8 & LPV \\
3 & 07:28:29.8 & 69:12:16 & 900\tablenotemark{c} & 16.89 & 0.99 & 0.7 & LPV \\
4 & 07:28:50.7 & 69:14:39 & -- & 17.51 & 0.60 & 0.8 & LBV? \\
5 & 07:28:49.4 & 69:14:01 & 400 & 17.71 & 0.43 & 0.7 & LPV \\
6 & 07:28:48.5 & 69:13:44 & 550 & 17.56 & 0.54 & 0.8 & LPV \\
7 & 07:28:46.6 & 69:13:25 & 700 & 17.01 & 0.72 & 0.9 & LPV \\
8 & 07:28:44.8 & 69:13:05 & 500 & 17.50 & 0.63 & 1.0 & LPV? EB?\\
9 & 07:29:02.5 & 69:14:35 & 550 & 17.36 & 0.81 & 1.0 & LPV \\
10 & 07:28:57.4 & 69:13:42 & 850\tablenotemark{c} & 16.58 & 0.30 & 0.6 & LPV \\
11 & 07:28:48.7 & 69:12:34 & 600 & 17.40 & 0.54 & 0.8 & LPV \\
12 & 07:28:49.2 & 69:12:22 & 600 & 17.60 & 0.54 & 0.7 & LPV? \\
13 & 07:28:42.6 & 69:11:32 & 850\tablenotemark{c} & 16.33 & 0.28 & 0.4 & LPV \\
14 & 07:28:59.8 & 69:13:32 & 550 & 17.64 & 0.69 & 0.7 & LPV \\
15 & 07:28:50.9 & 69:12:15 & 750 & 17.08 & 0.01 & 0.8 & LPV \\
16 & 07:28:49.1 & 69:11:54 & 850\tablenotemark{c} & 16.78 & 0.34 & 0.9 & LPV \\
17 & 07:29:02.3 & 69:13:16 & 500 & 17.67 & 0.48 & 0.8 & LPV \\
18 & 07:28:40.3 & 69:10:31 & 650 & 16.37 & 0.29 & 0.4 & LPV? \\
19 & 07:29:09.7 & 69:13:53 & 600 & 17.26 & 0.55 & 1.0 & LPV \\
20 & 07:28:58.9 & 69:12:31 & 700 & 17.09 & 0.76 & 0.8 & LPV \\
21 & 07:29:02.2 & 69:12:34 & 1300\tablenotemark{c} & 17.29 & 0.43 & 1.2 & LBV? \\
22 & 07:29:11.2 & 69:13:32 & 400 & 17.34 & 0.67 & 0.9 & LPV \\
23 & 07:28:52.1 & 69:11:13 & 650 & 17.53 & 0.23 & 0.8 & LPV \\
24 & 07:28:56.6 & 69:11:41 & 450 & 17.74 & 0.72 & 0.8 & LPV \\
25 & 07:28:48.2 & 69:10:35 & -- & 17.43 & 0.91 & 1.5 & LBV? \\
26 & 07:28:50.6 & 69:10:44 & 750 & 17.32 & 0.85 & 1.0 & LPV \\
27 & 07:29:08.7 & 69:12:15 & 400 & 17.41 & 0.52 & 0.7 & LPV \\
28 & 07:28:41.0 & 69:12:44 & 550 & 17.59 & 0.54 & 0.5 & LPV \\
29 & 07:28:49.5 & 69:12:25 & 650 & 17.61 & 0.89 & 0.5 & LPV \\
30 & 07:28:53.4 & 69:12:35 & 500 & 17.33 & 0.35 & 0.9 & LPV \\
31 & 07:28:59.0 & 69:14:15 & 500 & 17.65 & 0.61 & 1.1 & LPV \\
32 & 07:29:06.7 & 69:14:37 & 450 & 17.47 & 0.58 & 0.9 & LPV \\
\tableline
\end{tabular}
\end{center}
\caption{NGC 2366 Variable Stars}
\tablenotetext{a}{Measured from median images. The estimated uncertainties 
are $\pm 0.06$ magnitude for [3.6] and $\pm 0.03$ mag for [3.6]--[4.5].}
\tablenotetext{b}{Difference between the brightest and faintest points.}
\tablenotetext{c}{Complete cycle not observed.}
\end{table*}

\begin{deluxetable}{cccccccc}
\tabletypesize{\scriptsize}
\tablecaption{IC2754 Variable Stars}
\tablehead{DV\# & RA & Dec & Period & [3.6]\tablenotemark{a} & [3.6]--[4.5]\tablenotemark{a} & Amplitude\tablenotemark{b} & Type \\ 
 & (J2000) & (J2000) & (days) & (mag) & (mag) & (mag) & }
\startdata
1 & 10:28:57.0 & 68:24:13 & 650 & 17.00 & 0.45 & 0.4 & LPV \\
2 & 10:28:49.1 & 68:23:27 & 500 & 17.77 & 0.72 & 0.5 & LPV \\
3 & 10:28:51.5 & 68:24:27 & -- & 17.42 & 0.47 & 0.6 & LPV? \\
4 & 10:28:47.4 & 68:24:23 & 850 \tablenotemark{c}& 17.27 & 0.52 & 0.8 & LPV \\
5 & 10:28:49.1 & 68:25:05 & 500 & 17.63 & 0.69 & 1.0 & LPV \\
6 & 10:28:40.6 & 68:25:39 & 750 & 17.62 & 0.47 & 1.0 & LPV? \\
7 & 10:28:35.3 & 68:24:48 & 550 & 17.56 & 0.38 & 0.5 & LPV \\
8 & 10:28:42.1 & 68:26:15 & 850 \tablenotemark{c}& 17.00 & 0.12 & 0.8 & LPV \\
9 & 10:28:25.9 & 68:22:59 & -- & 16.76 & 0.17 & 1.0 & RSG? \\
10 & 10:28:41.1 & 68:26:22 & 500 & 17.05 & 0.46 & 0.7 & LPV \\
11 & 10:28:20.7 & 68:23:12 & 550 & 17.55 & 0.67 & 1.0 & LPV \\
12 & 10:28:21.2 & 68:23:26 & -- & 17.67 & 0.57 & 0.5 & LBV? \\
13 & 10:28:20.1 & 68:23:16 & -- & 17.57 & 0.56 & 0.5 & LBV? \\
14 & 10:28:30.7 & 68:25:38 & 650 & 17.62 & 0.70 & 0.8 & LPV? \\
15 & 10:28:31.4 & 68:25:53 & 1000 \tablenotemark{c} & 16.66 & 0.73 & 1.1 & sgB[e]? LPV?\\
16 & 10:28:21.6 & 68:23:55 & -- & 17.61 & 0.70 & 0.9 & LPV? \\
17 & 10:28:23.1 & 68:24:15 & 550 & 17.54 & 0.49 & 0.5 & LPV \\
18 & 10:28:22.7 & 68:24:29 & -- & 17.63 & 0.58 & 0.6 & LBV? \\
19 & 10:28:16.1 & 68:23:28 & 750 & 16.89 & 0.57 & 0.9 & LPV \\
20 & 10:28:21.3 & 68:24:35 & 700 & 17.53 & 0.32 & 0.6 & LPV \\
21 & 10:28:18.4 & 68:24:07 & 600 & 17.16 & 0.38 & 0.6 & LPV \\
22 & 10:28:14.8 & 68:23:27 & 550 & 17.19 & 0.48 & 1.1 & LPV? \\
23 & 10:28:12.6 & 68:23:06 & 550 & 17.10 & 0.67 & 1.0 & LPV \\
24 & 10:28:24.0 & 68:25:40 & -- & 17.48 & 0.51 & 0.7 & LBV? \\
25 & 10:28:03.8 & 68:22:42 & -- & 16.77 & 0.49 & 0.8 & LPV? \\
26 & 10:28:17.3 & 68:25:44 & 550 & 17.32 & 0.38 & 0.5 & LPV \\
27 & 10:28:14.2 & 68:25:19 & 450 & 17.62 & 0.52 & 0.6 & LPV \\
28 & 10:28:02.1 & 68:22:55 & -- & 15.53 & 0.62 & 0.3 & sgB[e]? \\
29 & 10:28:11.3 & 68:25:19 & -- & 16.43 & 0.32 & 1.0 & LBV? \\
30 & 10:28:03.4 & 68:24:13 & 850 \tablenotemark{c} & 17.16 & 0.06 & 0.7 & LPV \\
31 & 10:28:17.1 & 68:27:21 & 700 & 17.42 & 0.69 & 0.9 & LPV \\
32 & 10:28:08.0 & 68:26:58 & -- & 16.41 & 0.67 & 1.0 & sgB[e]? \\
33 & 10:27:55.2 & 68:25:08 & -- & 17.73 & 0.40 & 0.5 & LBV? \\
34 & 10:27:57.3 & 68:26:20 & 700 & 17.64 & 0.50 & 0.6 & LPV \\
35 & 10:28:16.3 & 68:25:04 & -- & 17.57 & --0.09 & 0.4 & LPV \\
36 & 10:28:09.8 & 68:25:03 & 600 & 17.62 & 0.68 & 1.0 & LPV \\
37 & 10:28:18.2 & 68:24:14 & 500 & 17.63 & 0.39 & 0.5 & LPV \\
\enddata
\tablenotetext{a}{Measured from median images. The estimated uncertainties 
are $\pm 0.06$ magnitude for [3.6] and $\pm 0.03$ mag for [3.6]--[4.5].}
\tablenotetext{b}{Difference between the brightest and faintest points.}
\tablenotetext{c}{Complete cycle not observed.}
\end{deluxetable}

	The periods listed in these tables should be viewed as preliminary only 
due to the spotty phase coverage in the light curves. 
For variables in which one complete cycle was not 
sampled then periods were estimated by assuming that the light curves are symmetric 
with time. The absence of a period entry in the tables indicates either no evidence of 
periodicity, and/or insufficient phase coverage. The limits in phase coverage, 
coupled with the finite time baseline, restricts the ability to measure periods to 
better than $\sim 10\%$. More specifically, the periods have estimated uncertainties on 
the order of 50 days if one or more cycles is observed, while those where a 
complete cycle was not observed have an estimated uncertainty on the order of 
100 days. Improved periods will result when the time baseline 
and the phase coverage are extended. These sources of uncertainty 
notwithstanding, it is demonstrated in Section 6 that the period estimates 
in the tables are sufficiently reliable to allow the PLRs of LPVs in 
each galaxy to be examined.

	The number of detected variables is smallest in Ho II. 
This probably does not reflect a difference in stellar 
content, as there is a larger inherent dispersion in the Ho II photometry when 
compared with the other galaxies, such that the $5\sigma$ detection criterion for the 
detection of variability results in an amplitude threshold that is larger in Ho II at 
a given magnitude than in the other galaxies. If the threshold for detecting variable stars 
in Ho II is lowered then more candidate variables would of course be 
detected, although contamination from non-variables would then become an issue. 

	The differential and cumulative histogram distributions of the
periods for LPVs and suspected LPVs are compared in Figure 7. The distributions 
have been normalized to the number of LPVs in each galaxy. The period distributions 
of NGC 2366 and IC 2574 have been shifted slightly along 
the horizontal axis to facilitate galaxy-to-galaxy comparisons. 

	The period distributions of NGC 2366 and IC 2574 are very similar. This 
agreement is perhaps not surprising given (1) the similarities in SFHs (Section 1), and 
(2) the homogeneous sampling cadence and temporal baselines of the data for each galaxy. 
As for Ho II, the smaller number of LPVs identified in that galaxy frustrate 
efforts to make meaningful comparisons with the other galaxies. There is 
a possible deficiency of LPVs in Ho II with periods near 600 days, although the 
significance of this is not clear given the uncertainties arising from small number 
statistics, coupled with the uncertainties in the period measurements. The mean 
periods of LPVs in all three galaxies agree within the errors in the mean. 

\begin{figure}
\figurenum{7}
\epsscale{1.0}
\plotone{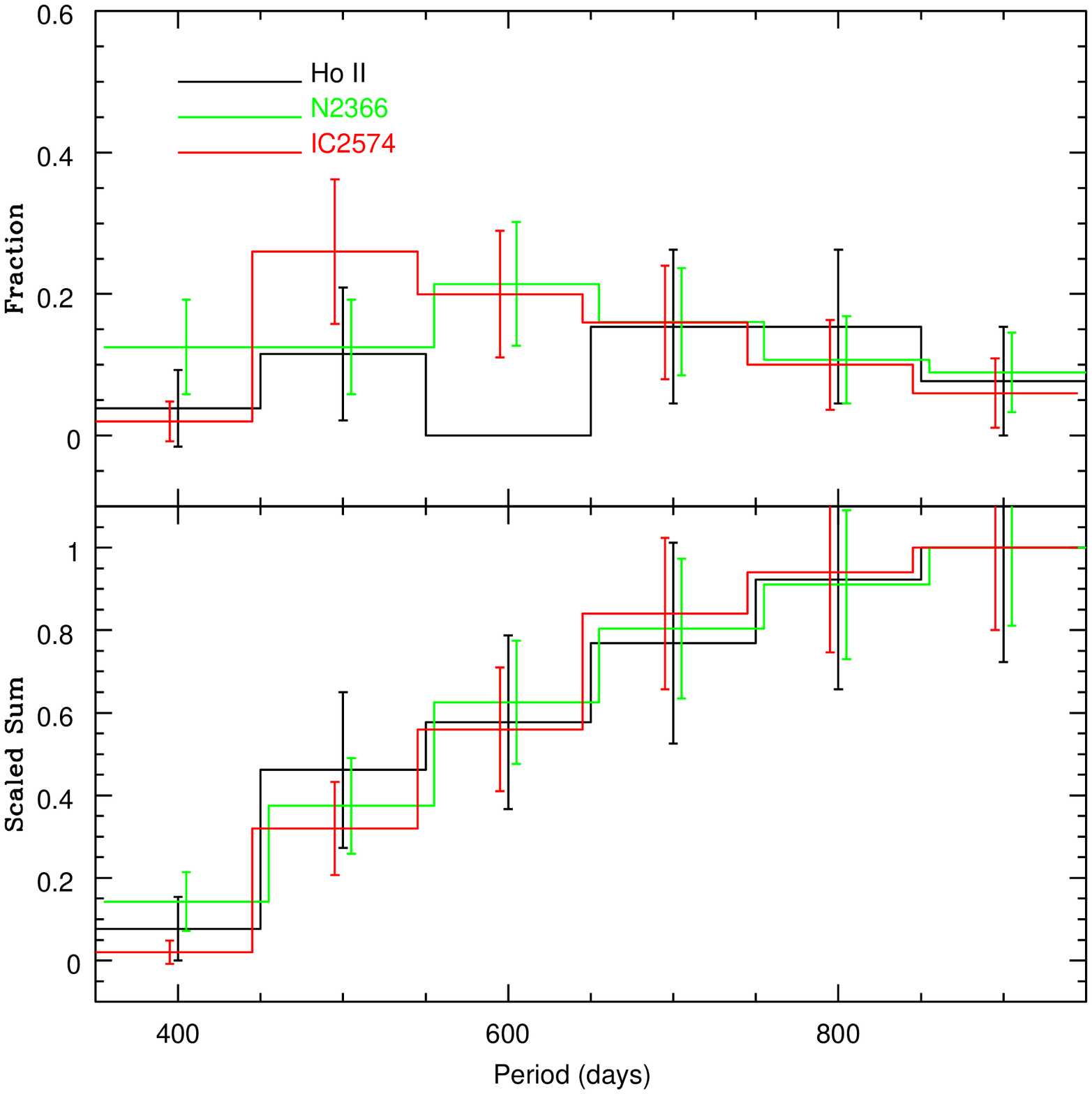}
\caption{Normalized differential (top panel) and cumulative (bottom panel) period 
distributions of LPVs and suspected LPVs. The NGC 2366 and IC 2574 distributions have been 
shifted slightly along the horizontal axis to facilitate galaxy-to-galaxy comparisons. The 
error bars show Poisson statistics, and do not include the uncertainties in the period 
measurements that smear the distributions along the horizontal 
axis. The period distributions of NGC 2366 and IC 2574 are 
in good agreement. While the period distribution of Ho II is compromised by 
small number statistics, there appears to be a deficiency of LPVs with periods near 600 days 
in that galaxy when compared with the others. The mean periods of LPVs in all three galaxies 
agree to within the formal error in their means.}
\end{figure}

	The referee has suggested that data from the Pan-STARRS (PS) Data Release 
2 (Chambers et al. 2019) be examined to determine if the variables can be detected 
at shorter wavelengths. Extending the wavelength coverage 
and time baseline of these objects are of obvious interest to better 
understand their properties. However, none of the LPVs were detected in the 
stacked PS images, let alone the individual images that would be required to construct 
light curves. This is perhaps to be expected, as the PS survey in
this part of the sky has a 50\% completeness fraction near $i' \sim 22$, which 
is near the expected peak brightness of the brightest LPVs at this wavelength. 
In any event, the [3.6]--[4.5] colors of the LPVs (see next section) indicate that 
the objects found in the IRAC images are surrounded by circumstellar shells, 
and hence may be heavily obscured at visible wavelengths. This is consistent 
with the majority of these being extreme AGB stars, that are near the 
end point of their AGB evolution (next section). As for the 
variables that are suspected massive stars, that these were also found in the IR 
means that there is a similar bias towards large amounts of obscuration. 
Some of suspected massive stars that have the smallest [3.6]--[4.5] colors, and 
hence are the least obscured, might be detected in the PS images, but the significance of 
these detections is clouded by the greater degree of source confusion that occurs at 
shorter wavelengths, coupled with the crowded enviroments in 
which these objects tend to be found.

\subsection{CMDs, Luminosity Functions, and Classifications}

	The $([4.5], [3.6]-[4.5])$ CMDs of the galaxies are shown 
in Figures 8, 9, and 10. The locations of variables in each CMD are indicated by 
filled (`LPV') and unfilled (`LPV?' and all other types) green squares. The 
vertical sequence near [3.6]--[4.5] $\sim 0$ at the bright end of the CMDs is populated by 
foreground stars and background galaxies. The vertical nature of this sequence reflects 
the low temperature sensitivity of the [3.6]--[4.5] color 
among objects with photospheric temperatures (i.e. T$_{eff} > 2000$K). 

\begin{figure}
\figurenum{8}
\epsscale{0.55}
\plotone{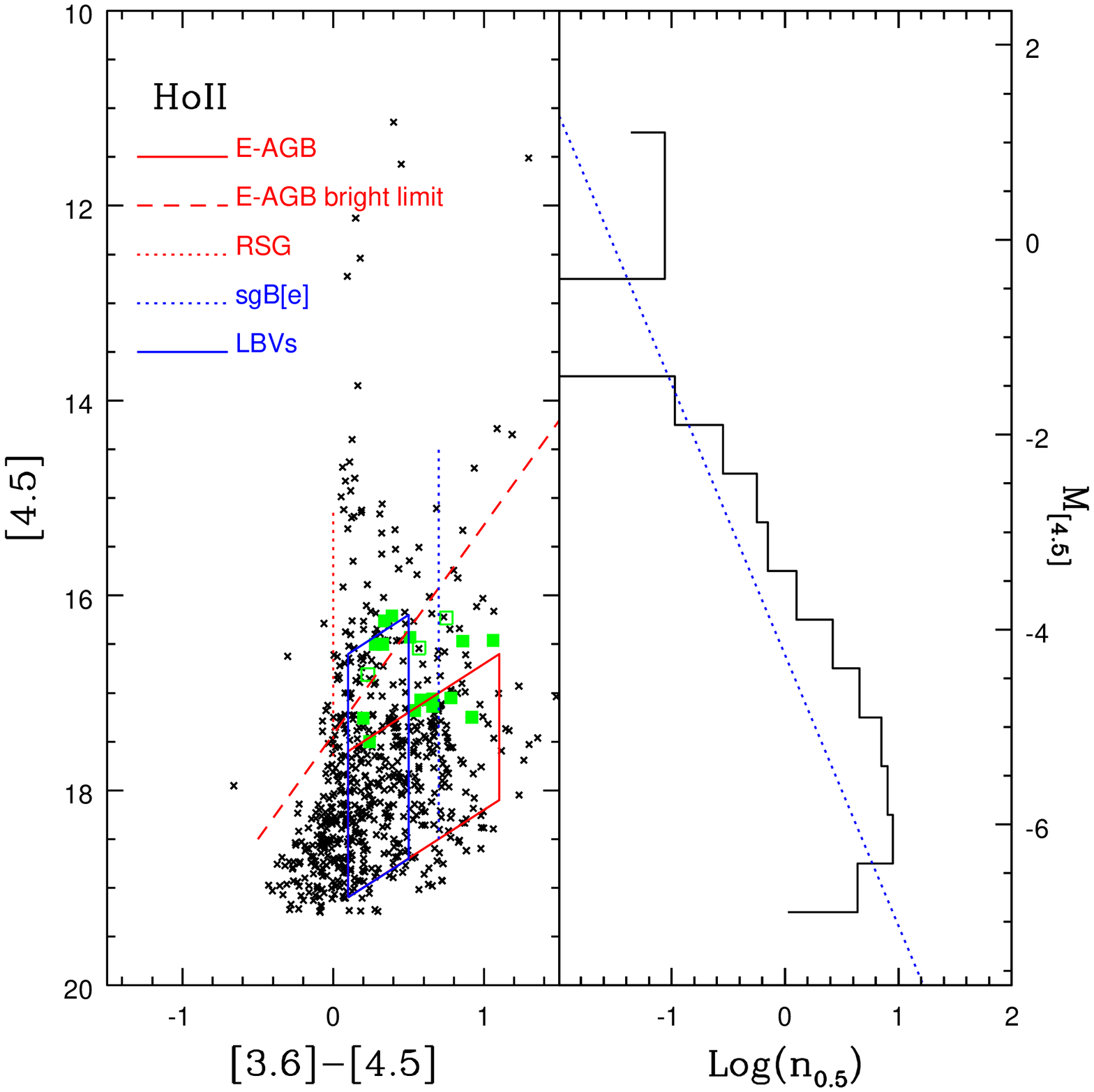}
\caption{Left hand panel: $([4.5],[3.6]-[4.5])$ CMD of Ho II. 
Confirmed LPVs are shown as filled green squares, and the [3.6]-[4.5] colors of many 
of these indicate that there is significant thermal emission from circumstellar 
material. The unfilled green squares mark variables that, based on their light 
curves and photometric properties, are either not certain LPVs, or are clearly not LPVs. 
The dashed red line shows the upper envelope of E-AGB stars 
in the LMC, taken from Figure 1 of Boyer et al. (2015) and adjusted to the Ho II 
distance modulus listed in Table 1, while the red box indicates the location of the 
majority of LMC E-AGB stars in Figure 1 of Boyer et al. (2015). 
The dotted red line shows the approximate locus of LMC RSGs 
in Figure 2 of Bonanos et al. (2009) shifted to the distance of Ho II, 
while the dashed blue line shows the locus of LMC sgB[e] stars and the blue 
box indicates the region occupied by bright LMC LBVs in Figure 2 of Bonanos 
et al. (2009). The LPVs in Ho II that are found in the IRAC images have 
luminosities that are comparable to or slightly brighter than 
the vast majority of E-AGB stars in the LMC. There are sources in the CMD above the 
E-AGB limit that are possible massive stars with hot circumstellar envelopes. Right 
hand panel: [4.5]--band LF of all sources in the Ho II image, where $n_{0.5}$ is the 
number of sources per 0.5 magnitude interval per square arcmin. The absolute 
magnitude scale along the right hand axis uses the Ho II distance modulus 
listed in Table 1. The dotted blue line is a least squares fit 
of a power law relation to source counts near the edges of the three galaxy 
fields in Figure 1. The source counts in Ho II systematically exceed 
those indicated by the blue line when [4.5] $> 14.5$, and it is suggested that there is 
a population of very luminous RSGs in Ho II that is not seen in the other galaxies.}
\end{figure}

\begin{figure}
\figurenum{9}
\epsscale{1.0}
\plotone{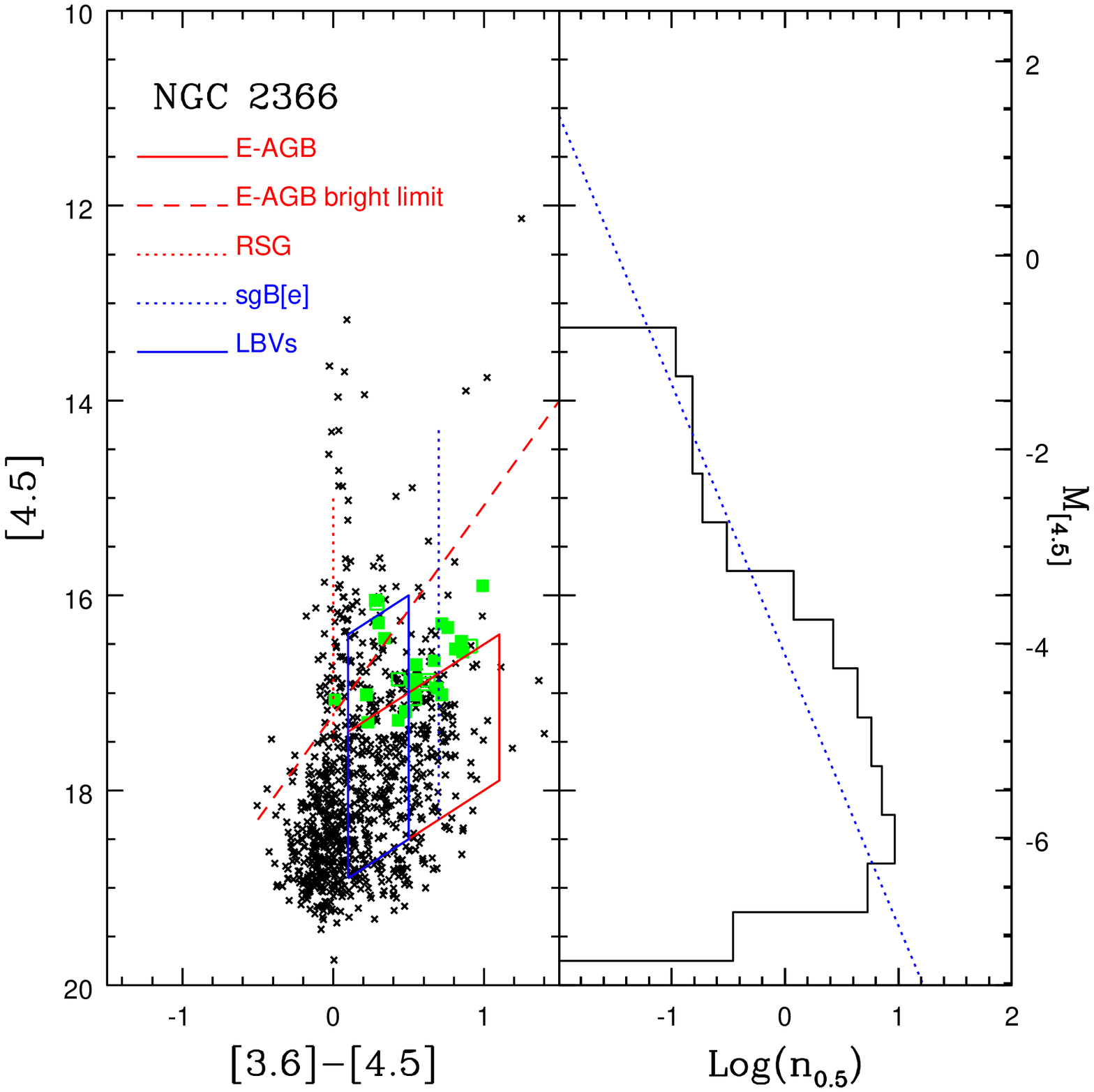}
\caption{Same as Figure 8, but for NGC 2366. The placement of the fiducial LMC sequences 
assumes the distance modulus for NGC 2366 listed in Table 1. The sudden onset of 
stars in the CMD and LF near [4.5] = 15.75 contrasts with the LF of Ho II, which does 
not have a corresponding discontinuity. There are sources in 
NGC 2366 that are near the fiducial sgB[e] sequence, as well as 
near the bright end of the area occupied by LBVs.}
\end{figure}

\begin{figure}
\figurenum{10}
\epsscale{1.0}
\plotone{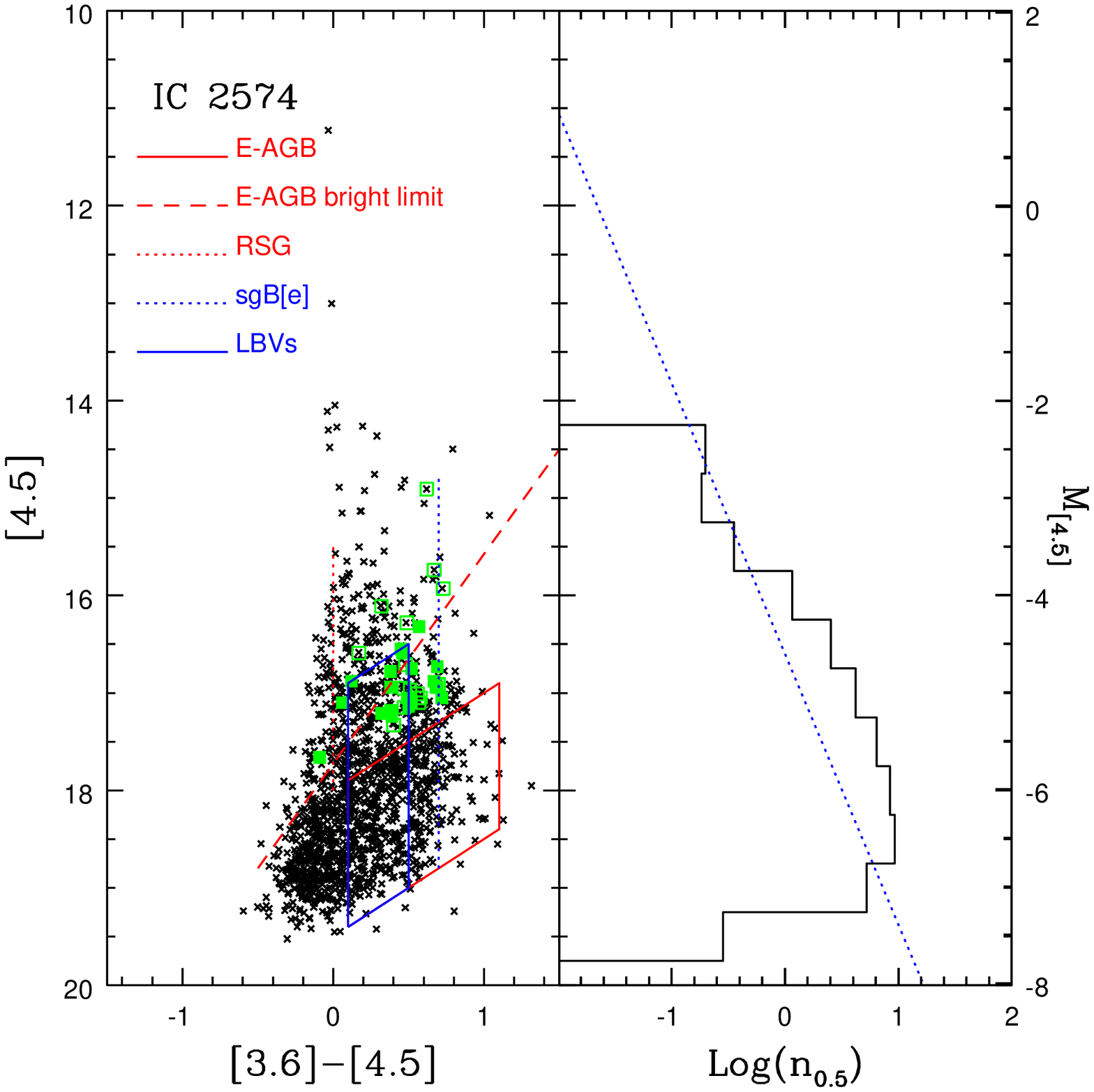}
\caption{Same as Figure 8, but for IC 2574. The placement of the LMC 
sequences assume the IC 2574 distance modulus listed in Table 1. As with 
the other galaxies, the LPVs found in IC 2574 are 
among the most luminous E-AGB stars. There are also a number of candidate 
massive stars, some of which are variables.}
\end{figure}

	The CMDs broaden along the color axis near [4.5] $\sim 15 - 16$, due to 
the onset of luminous, highly evolved stars that are encased in warm circumstellar dust 
shells. Contamination from foreground stars and unresolved background galaxies remains 
significant in that magnitude range, especially among objects with 
[3.6] -- [4.5] $\sim 0$. To gauge the extent of this contamination, number counts 
were made in regions that are near the edges of each of the galaxy images in Figure 1. 
As the three galaxies have similar Galactic latitudes, then
the background counts from all three galaxies were 
combined to reduce statistical uncertainties.

	The LF of all sources in each galaxy are plotted in the right hand panels 
of Figures 8, 9, and 10, where n$_{0.5}$ is the number of objects per 
square arcminute per 0.5 magnitude interval in [4.5]. 
The dotted blue lines in these panels are a power-law that 
was fit to the number counts in the background/foreground 
areas of all three fields. Fitting a relation to the combined background/foreground counts 
suppresses the uncertainties in the measurements in individual magnitude bins.
The fitted relation matches the number of objects at intermediate magnitudes in the 
LFs of NGC 2366 and IC 2574, as expected if it tracks sources that do not belong to the 
target galaxies.

	There is a break in number counts near [4.5] $\sim 16$ in 
the LFs of NGC 2366 and IC 2574, and the LFs of those galaxies 
exceed the expected number of non-galaxy members
at magnitudes fainter than these breaks. The breaks in the LFs 
occur at magnitudes that correspond to the onset of large numbers of stars with 
[3.6]--[4.5] $> 0$ in the CMDs of NGC 2366 and IC 2574, 
as expected if the discontinuities in the LFs track the 
onset of highly evolved stars in each galaxy. In contrast, the LF of Ho II follows a 
more-or-less continous trend over a wide range of magnitudes with no 
obvious break. There is also an excess number of sources when compared with 
the background/foreground counts up to $\sim 1.5$ magnitudes brighter than in the other 
galaxies; stars in Ho II thus appear to occur in statistically significant numbers 
to as bright as [4.5] $\sim 14.5$. These results likely indicate that 
there is a population of very bright RSGs in Ho II that is not present in the 
other galaxies. Indeed, RSGs in the LMC have [3.6]--[4.5] $\sim 0$, and 
the most luminous RSGs in the LMC would occur near [4.5] $\sim 14.5$ 
if shifted to the distance of Ho II (Bonanos et al. 2009), matching the peak 
of the Ho II sequence. As the [3.6]--[4.5] colors of RSGs are similar to those of 
foreground stars then a population of RSGs would not cause an obvious widening of the 
CMD along the color axis. 

	The photometric faint limit varies with location throughout each field, 
due to the obvious non-uniform distribution of fainter stars that make up these galaxies. 
However, there is evidence that the faint limit of the photometry tends not to change 
by more than $\sim 0.5$ magnitude in [4.5] with location in each image. 
The magnitude at which sample incompleteness sets 
in can be approximated by the inflexion point in the LF where the number counts 
turn down, and this occurs near [4.5] $\sim 18.5$ in the LFs. While not shown in 
Figures 8 -- 10, the inflexion point in the foreground/background fields occurs 
near [4.5] $\sim 19$, which is only $\sim 0.5$ magnitude fainter than in the galaxy LF. 
This suggests that the stellar samples in each galaxy are likely 
complete to [4.5] = 18.5. We caution that this does not mean that
the sample of LPVs is complete at these magnitudes, as the procedure to 
detect variables relies on an amplitude criterion, and 
LPVs with small amplitude light variations are missed.
That the photometric completeness limit does not vary greatly with stellar density at 
these wavelengths is not surprising, as the stars in Figures 
8 -- 10 are among the brightest objects in these galaxies at these wavelengths, 
and far exceed in brightness the unresolved stars that make up the bulk of the stellar 
content. Hence, they are not as susceptible to the crowding issues that occur 
among stars at fainter magnitudes and shorter wavelengths.

\subsubsection{Candidate LPVs}

	The LMC contains a well-studied population of AGB stars that is 
a useful reference for assessing the stellar contents of more distant galaxies. An extensive 
database of measurements of LMC and SMC AGB stars was obtained as part of 
the SAGE survey (e.g. Blum et al. 2006). This survey identified a number of 
stars that are nearing the end of their AGB evolution and show signs of significant 
circumstellar extinction. These extreme AGB (E-AGB) stars have high 
intrinsic luminosities and [3.6]--[4.5] colors that place them to the right of the 
dominant stellar plume on the $([4.5], [3.6]-[4.5])$ CMD. The red [3.6]--[4.5] color of 
these objects is due to emission from circumstellar dust that has accumulated as the 
stars shed mass during the latter stages of evolution. 
Blum et al. (2006) find that the majority of E-AGB stars in the LMC have 
[3.6]--[4.5] between 0.1 and 1.5 magnitudes, and M$_{[4.5]}$ between --8 and --11 
magnitudes.

	Boyer et al. (2015) examine the IR properties of luminous dust-obscured stars 
in nearby dwarf galaxies, using the Blum et al. (2006) photometry as a reference. 
The area in their $([3.6], [3.6]-[4.5])$ CMDs that contains 
the main body of E-AGB stars in Figure 1 of Boyer et al. (2015) 
has been translated into the $([4.5], [3.6]-[4.5])$ plane, and the result is shown in 
Figures 8 -- 10 after adjusting for the distances to the M81 group 
galaxies. The dashed line marks the approximate upper envelope of 
E-AGB stars in Figure 1 of Boyer et al. (2015). 

	The majority of LPVs in the three galaxies populate the area of the 
CMD that contains the most luminous E-AGB stars. 
Spectroscopic and photometric follow-up will likely reveal that 
many of these are C stars, as such objects are expected to be 
common among highly evolved AGB stars in galaxies that have subsolar 
metallicities and large intermediate age populations. That many of the LPVs are near the 
upper IR brightness limit for E-AGB stars is a selection effect that results from 
the faint limit of the SPIRITS observations, coupled with the distances of the 
galaxies and the magnitude interval searched for LPVs. The dominant clump of LPVs in the 
CMD of NGC 2366 falls along a locus of roughly constant [3.6], accentuating the [3.6] 
magnitude selection limit. A sizeable population of fainter LPVs awaits detection in 
each galaxy. 

\subsubsection{Candidate Massive Stars}

	There is a spray of objects that extends $\sim 1 - 2$ magnitudes above the 
E-AGB cut-off in the CMDs, and thus occupies a region of the CMD that is home to 
massive stars. Bonanos et al. (2009) discuss the IR properties of massive 
stars in the LMC, with red supergiants (RSGs) among the brightest of these. 
The RSGs in the Bonanos et al. (2009) sample tend to have [3.6]--[4.5] colors 
between --0.2 and 0.2 magnitudes, and can be as bright as [3.6] $\sim 6$ magnitudes in the 
LMC, or M$_{[3.6]} \sim -12.5$ magnitudes. This corresponds to [4.5] $\sim 15$ magnitudes 
in Ho II and NGC 2366 if [3.6] -- [4.5] $\sim 0$ magnitudes, and [4.5] $\sim 15.5$ magnitudes 
in IC 2574. As noted earlier, the expected peak brightness for RSGs more-or-less 
corresponds to the magnitude where the number counts in Ho II exceed those expected 
solely from background/foreground objects. An excess in number counts is not seen in 
the LFs of the other two galaxies at similar magnitudes, suggesting that Ho II may have a 
population of luminous RSGs that is not present in the other galaxies.

	Some of the variables found here are probably RSGs. 
DV10 in Ho II is classified as a possible RSG in Table 3, 
although it also falls in the region of the CMD that is occupied by LBVs. 
The light curve of DV10 shows an initial dimming episode followed by brightening, and 
these variations span $\sim 0.7$ magnitude in [3.6]. There is also 
$\pm 0.2$ magnitude scatter about the longer term trends. DV9 in IC 2574 is another 
possible RSG. The light curve of that star shows an overall dimming trend, although 
there is a discrepant point that suggests more complex light variations are present. 
Both of these variables fall near the upper boundary of the area in the CMDs that is 
occupied by LBVs, and so there is ambiguity in their classification. 

	Some of the candidate massive stars may be LBVs. Bonanos et al. (2009) find 
LBVs as bright as M$_{[3.6]} \sim -11$ magnitudes in the LMC, which corresponds to 
[4.5] $\sim 16.5$ magnitudes in Ho II and NGC 2366 if [3.6] -- [4.5] $\sim 0$ magnitudes, 
and [4.5] = 17 magnitudes in IC 2574. LBVs in the Bonanos et al. (2009) sample span almost 
3 magnitudes in [4.5], and the area of the $([4.5], [3.6]-[4.5])$ CMD occupied 
by the LBVs in Figure 3 of Bonanos et al. (2009), 
adjusted for the distances to the target galaxies, 
is indicated with a blue box in Figures 8, 9, and 10.

	There is considerable overlap between the areas occupied by LBVs and E-AGB 
stars in these CMDs, with the upper boundary of the region occupied by LBVs 
only a few tenths of a magnitude in [4.5] above the upper envelope of E-AGB stars. 
In fact, LPVs are identified near the upper boundary of the LBV region on 
the CMD, hinting at substantial contamination from E-AGB stars over the 
area on the CMD that contains LBVs. Lacking spectra, the primary means of isolating 
candidate LBVs with these data is then through variability. There are a handful of 
variables in the galaxies that fall within the LBV area of the CMD that do not show 
light curves indicative of LPVs, and there are three 
such objects in IC 2574. The light curve of DV9 in IC 2574
shows evidence of semi-regular variations, although it is listed as a possible RSG 
variable in Table 5 based on its [3.6]--[4.5] color. 
As for the other candidate LBVs in IC 2574, the light 
curve of DV25 shows a progressive decrease in [3.6] magnitudes 
when these data were recorded, while that of DV29 shows a progressive 
increase in brightness.

	Despite having hot photospheric temperatures, supergiant B[e] (sgB[e]) stars may 
be among the brightest IR sources in a star forming galaxy. The IR 
emission from these objects originates in circumstellar shells, and so they 
are potentially challenging targets for spectroscopy at visible and red 
wavelengths if the shell is optically thick. Bonanos et al. (2009) 
find that sgB[e] stars are as bright as [3.6] = 6 magnitudes in the LMC (i.e. 
as bright as RSGs in that passband), and tend to have [3.6]--[4.5] $\sim 0.7$ 
magnitudes. Therefore, the most luminous sgB[e] stars may be many tenths of a magnitude 
brighter than the brightest RSGs in the $([4.5], [3.6]-[4.5])$ CMD.

	The locus of LMC sgB[e] stars is shown as a 
dashed blue line in the left hand panels of Figures 8, 9, and 10. 
Each of the galaxies contain objects that fall close to the sgB[e] 
sequence at [4.5] magnitudes that are well above the E-AGB locus. IC 2574 contains the 
largest number of very luminous candidate sgB[e] stars, and one of these (DV28) shows 
evidence of variability that is not indicative of a LPV. The light curve of DV28 
gradually increases in [3.6] during the time interval sampled with these data, 
and its location on the CMD suggests that it is one of the most 
luminous objects in IC 2574. DV15 and DV32 in IC 2574 are also near the sgB[e] 
sequence. The light curve of the former shows a steady $\sim 0.5$ magnitude increase 
in [3.6] during the time that these data were recorded. 
There is no evidence of the intrinsic scatter that appears in some of the light curves 
of other candidate massive stars. In contrast, the light curve of DV32 shows evidence 
of $\pm 0.5$ magnitude dispersion in [3.6], with a possible overall dimming trend.

	Unlike RSGs and LBVs, candidate sgB[e] stars occur in a part of the $([4.5], 
[3.6]-[4.5])$ CMD that is not susceptible to contamination from foreground or E-AGB 
stars, and so a sample of these objects can be identified without light curve information.
The locations, [4.5] magnitudes, and [3.6]--[4.5] colors of non-variable stars 
that fall close to the sgB[e] sequence in the CMDs of the three galaxies and 
that are more than 0.5 magnitudes above the E-AGB upper locus in the LBV area 
on the CMD are listed in Table 6. With the caveat that the number of objects is small, 
IC 2574 contains twice as many sgB[e] candidates as in the other two galaxies combined. 
This could suggest a more active SFR in IC 2574 during the past few Myr, or perhaps 
characteristics of the ISM in IC 2574 that are more amenable to the formation 
of very massive stars (e.g. Kirk \& Meyers 2012).

\begin{table*}[ht!]
\begin{center}
\begin{tabular}{rcccc}
\tableline\tableline
\# & RA & Dec & [3.6]\tablenotemark{a} & [3.6]--[4.5]\tablenotemark{a} \\
 & (J2000) & (J2000) & (mag) & (mag) \\
\tableline
Ho II B1 & 08:19:29.0 & 70:42:58 & 15.792 & 0.685 \\
 & & & & \\
NGC 2366 B1 & 07:28:45.9  & 69:11:25 & 14.985 & 0.417 \\
B2 & 07:28:58.0  & 69:11:33 & 14.897 & 0.526 \\
 & & & & \\
IC 2574 B1 & 10:28:14.3  & 68:25:57 & 15.842 & 0.663 \\
B2 & 10:28:35.9  & 68:23:50 & 15.838 & 0.603 \\
B3 & 10:28:29.5  & 68:24:03 & 15.816 & 0.684 \\
B4 & 10:28:03.1  & 68:24:33 & 15.060 & 0.605 \\
B5  & 10:28:47.1  & 68:22:32 & 15.612 & 0.710 \\
B6 & 10:27:48.3  & 68:24:42 & 15.184 & 1.040 \\
 & & & & \\
\tableline
\end{tabular}
\end{center}
\caption{Non-variable candidate sgB[e] Stars}
\tablenotetext{a}{Measured from median images.}
\end{table*}

\section{SPATIAL DISTRIBUTIONS}

	The on-sky locations of the variable stars track 
areas of star formation during recent epochs. The variables that are suspected 
massive stars likely have ages $\leq 10 - 20$ Myr. As for LPVs, 
those objects with the longest periods in moderately metal-poor environments 
may have progenitor masses in excess of 4 M$_{\odot}$ (Whitelock 
et al. 2018), and hence ages of no more than $\sim 100$ Myr (e.g. Cordier 
et al. 2007). LPVs with periods near 400 days may include 
the least massive -- and hence oldest -- progenitor stars. While these may have 
near-solar progenitor masses (Takeuti et al. 2013), there is considerable scatter in any 
period-mass relation, and at least some of the shortest period LPVs found in this 
study are likely overtone pulsators (Section 6).

	The on-sky locations of LPVs and candidate LBVs and sgB[e] stars are compared 
with those of non-variable stars and unresolved light in Figure 11. 
The non-variable stars are restricted to [4.5] magnitudes 
between 16 and 18 to maintain photometric completeness. The images of unresolved 
light are those of background light that were constructed by smoothing 
star-subtracted images of the galaxies (Section 3).
Foreground stars and compact background galaxies are a complicating factor 
when examining the distribution of non-variables, as the predicted 
counts for these objects in Figures 8 -- 10 indicate that they can account 
for up to 50\% of the objects at the magnitudes examined here (Section 4). These 
contaminants are expected to be more-or-less uniformly distributed on the sky. 

\begin{figure}
\figurenum{11}
\plotone{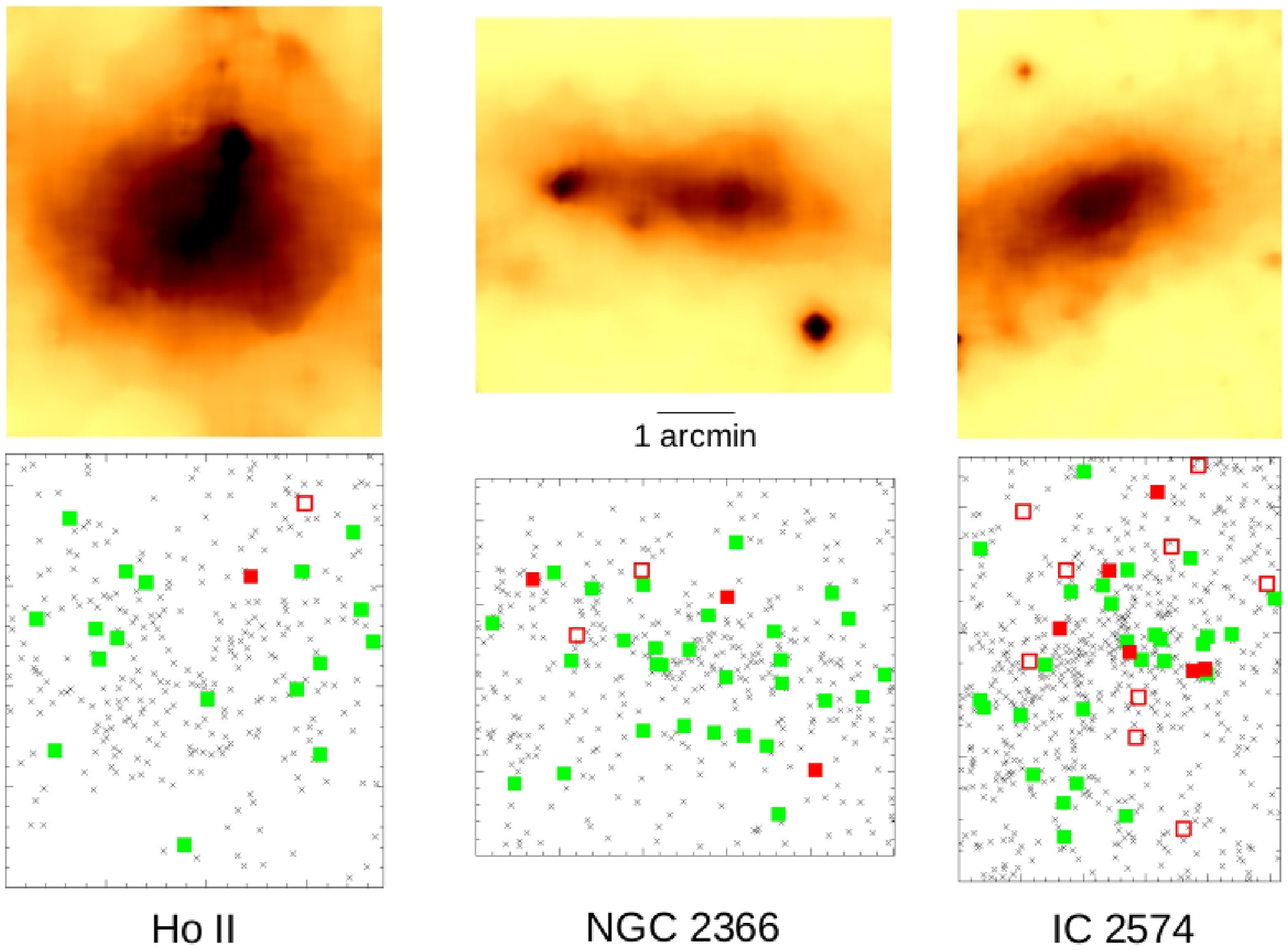}
\caption{Comparing light from unresolved stars (top row), with the spatial distributions 
of non-variable stars with [4.5] between 16 and 18 (lower row, gray crosses). 
Also highlighted in the lower panels are the locations of LPVs (green squares), 
candidate LBVs (filled red squares), and sgB[e] stars 
(unfilled red squares). The images in the top row are the background light templates 
discussed in Section 3. LPVs and candidate massive stars in Ho II 
tend to avoid the area of maximum stellar density, and there are 
gaps in the distribution of non-variable stars. As with Ho II, LPVs and candidate 
massive stars in NGC 2366 and IC 2574 are more uniformly distributed than 
non-variable stars, although non-variable stars in IC 2574 have a 
tendency to concentrate in areas of higher surface brightness than is the 
case in Ho II.}
\end{figure}

	The LPVs with the longest periods in these galaxies are likely near 
their birth places. Evidence to support this claim is presented by Bastian et al. (2011), 
who examined the two point correlation function of stars in various age intervals to 
estimate the disruption timescale of stellar groupings in these galaxies. 
They conclude that groupings in Ho II survive for up to 225 Myr, while in IC 
2574 they last 150 - 200 Myr, and in NGC 2366 100 Myr. 

	Small number statistics are an obvious issue when assessing the distribution of 
LPVs and candidate massive stars in Ho II. Keeping this caveat in mind, 
the distribution of LPVs in Ho II appears not to follow the underlying IR light, 
in the sense that the LPVs are more diffusely distributed. 
The diffuse distribution of the LPVs in Ho II is consistent with them 
forming well after the bulk of RGB and fainter AGB stars that presumably 
dominate the unresolved light. The sole candidate sgB[e] star in Ho II is offset from the 
main body of the galaxy, and is part of the star forming complex to the East of the 
galaxy center.

	Bernard et al. (2012) found that the exponential scale length 
of AGB and RGB stars in Ho II is 0.76 - 0.80 arcmin, whereas bright 
blue MS and blue supergiant stars have a scale length in excess 
of 3 arcmin. The projected distribution of LPVs and other bright AGB stars found in the 
IRAC images of Ho II differ from the distribution of AGB stars 
studied by Bernard et al. (2012) because the two AGB samples 
have different mean ages. While the area of the CMD that Bernard et al. (2012) use to 
identify AGB stars almost certainly contains some young AGB stars, the dominant 
component consists of stars that are immediately above the RGB in the CMD 
(Figure 4 of Bernard et al. 2012). The Bernard et al. sample thus 
contains older populations, rather than the intrinsically 
luminous AGB stars that have been detected in the 
IRAC images. A further complication for the detection of 
young LPVs at the wavelengths considered by Bernard et al. (2012) 
is that E-AGB stars can have substantial circumstellar extinction.

	The non-variable star sample in Ho II includes fainter objects than the LPVs, and 
so likely contains stars that span a broader range of progenitor masses, and hence ages.
Still, the distribution of the non-variable stars in Ho II does not appear to track 
surface brightness in the [3.6] background image, in the sense that stars are not 
detected near the galaxy center. This is part of a general 
trend throughout Ho II: non-variable AGB stars appear not 
to track surface brightness, even near the edges of 
the galaxy. Gaps are also seen in the distribution of non-variable stars in 
Ho II. The locations of these holes do not correlate with the HI bubbles found by 
Puche et al. (1992), although a possible exception is the gap 
in the stellar distribution near the center of Ho II that coincides with bubble 21 in the 
Puche et al. (1992) study. This is the largest HI bubble in Ho II.

	The projected distribution of LPVs in NGC 2366 and IC 2574 are similar to that 
in Ho II, in the sense that the LPVs in these galaxies tend to be distributed 
more-or-less uniformly across the field, with no obvious correlation between number 
density and [3.6] surface brightness. The candidate massive stars in IC 2574 are much more 
widely distributed than the unresolved light from these galaxies, although 
the distribution of non-variable E-AGB stars in IC 2574 more closely tracks 
unresolved light than in Ho II, in the sense that there is a clear concentration of E-AGB 
stars along the major axis of IC 2574. The orientation of 
both galaxies makes it difficult to search for gaps in the distribution of E-AGB 
stars like those in Ho II.

\section{PERIOD-LUMINOSITY RELATIONS AND DISTANCES}

	The PLRs of LPVs and suspected LPVs in the three galaxies 
are compared in Figure 12. LPVs occupy multiple sequences on the 
period-luminsity plane (e.g. Wood et al. 1999; Riebel et al. 2010). 
Following the nomenclature described by Wood et al. (1999), the majority of LPVs are 
expected to fall along Sequence C, which is populated by fundamental 
mode pulsators. First overtone pulsators fall along Sequence C' 
$\sim 1$ magnitude above Sequence C on the period $vs$ [3.6] 
magnitude plane. Roughly one third of LPVs fall along Sequence D, 
which is populated by long secondary period stars (e.g. Riebel et al. 2010). 
Sequence D is located a few magnitudes below Sequence C in the PLRs, and so
is near the photometric faint limit of this study. 
While no stars on Sequence D are detected in this study, some LPVs 
found here fall midway between Sequences C and D (see below). 

\begin{figure}
\figurenum{12}
\epsscale{0.8}
\plotone{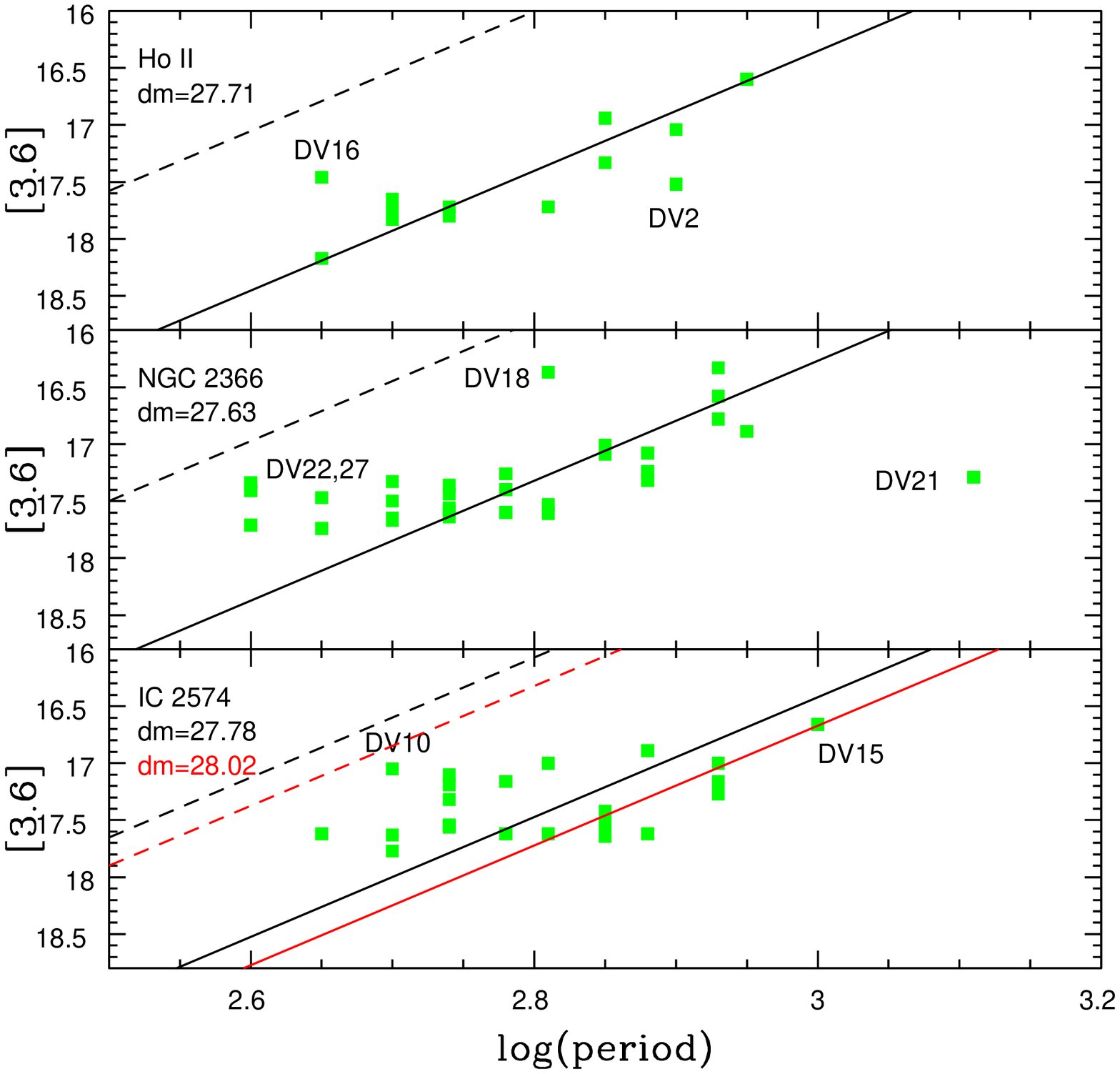}
\caption{[3.6]-band PLRs, with the period in days. The black line in each panel is 
Equation 1 of Goldman et al. (2019), shifted by the 
distance modulus specified in each panel, while the dashed line 
shows the expected location of Sequence C'. Variables that 
depart from the fundamental mode sequence and are discussed in the 
text are labelled. The red line in the lower panel shows the Goldman et al. (2019) 
PLR if the RGB-tip distance modulus of IC 2574 listed in Table 1 is adopted. 
The apparent PLRs of NGC 2366 and IC 2574 are flatter than the Goldman 
et al. (2019) relation, although there is reasonable agreement with the Goldman et al. 
(2019) relation at longer periods. It is concluded that the short period end of the PLR in NGC 
2366 and IC 2574 is skewed by the photometric faint limit and overtone pulsators.}
\end{figure}

	Goldman et al. (2019) examine the [3.6] and [4.5] PLRs of LPVs in galaxies 
that span a range of metallicities. There is a 0.25 magnitude 
scatter in the [3.6] relation, with no evidence 
of a metallicity dependance. The [3.6] PLR from Goldman et al. 
(2019), shifted by the distance modulus listed in each panel, is compared with the 
observed PLRs in Figure 12. The comparisons in Figure 12 indicate that 
the PLRs of the three galaxies have differing {\it apparent} slopes in the 
sense that the relations for NGC 2366 and IC 2574 are flatter than the Goldman et 
al. (2019) relation. The source of these differences is discussed below.

	The LPVs in Ho II fall on or close to Sequence C, and the majority of these define 
a relation that is well matched by that defined by Goldman et al. (2019). 
An exception is DV16, which falls roughly midway between the fundamental and overtone 
pulsation sequences. Another LPV that departs from the PLR is DV2, which 
is a few tenths of a magnitude below the trend defined by the other LPVs. 
DV2 has the reddest [3.6]--[4.5] color of the Ho II LPVs, opening the possibility 
that its intrinsic [3.6] brightness may be affected by circumstellar extinction. 
The tight relation defined by these data is evidence that there are not 
significant uncertainties in the estimated periods. If all points are 
considered when matching the Goldman et al. PLR then a distance modulus of 
$27.71 \pm 0.10$ is found, with a dispersion of $\pm 0.33$ mag. The uncertainty 
in the distance modulus is the $1\sigma$ formal error in the mean, and does not include the 
uncertainties in Equation 1 of Goldman et al. (2019). This distance modulus is in excellent 
agreement with the RGB-tip based distance modulus of Ho II listed in Table 1.

	LPVs in NGC 2366 define a broader scatter envelope in the PLR 
than those in Ho II. DV21 has the longest period measured among variables in NGC 2366, 
and it is classified as a possible LBV in Table 4. It is included in Figure 12 to 
examine the possibility that it might be an LPV. It is located 
almost 1.5 magnitudes below the fitted relation, placing it midway 
between Sequences C and D. A handful of LPVs in a similar 
part of the PLR are also seen in Figure 6 of Goldman et al (2019), 
where the PLR of LPVs in a large number of nearby metal-poor galaxies is shown.
Unlike DV2 in Ho II, DV21 has a median [3.6]--[4.5] color that 
is near the midpoint of the other LPVs, suggesting that it is not subject to 
exceptionally high circumstellar extinction when compared with other 
LPVs in this sample. We suspect that DV21 is likely not an LPV.

	A blind application of the Goldman et al. (2019) calibration 
to the entire sample of NGC 2366 LPVs yields 
a distance modulus of $27.52 \pm 0.09$, with a dispersion $\pm 0.50$ magnitudes. 
However, a comparison with the Goldman et al. (2019) relation shifted to this 
distance modulus found that there are LPVs other than DV21 in the NGC 2366 PLR that 
depart from the relation defined by all objects. In fact, 
there is a systematic trend, in the sense that LPVs with periods $\leq 500$ 
days consistently fall above the fitted Goldman et al. (2019) relation.

	The solid line in the middle panel of Figure 12 shows the Goldman et al. 
relation based on the distance modulus defined by LPVs with periods longer 
than 500 days, and excluding DV21. The distance modulus 
defined by these stars is $27.63 \pm 0.07$, with a dispersion about the 
relation of $\pm 0.30$ magnitudes. The dashed line shows the expected sequence C' with 
that distance modulus. DV18, DV22, and DV27 fall within 
a few tenths of a magnitude of Sequence C', making them 
possible first overtone pulsators, although the 650 day period for DV18 
is long for an overtone pulsator.

	The PLR for IC 2574 is shown in the bottom panel of Figure 12. DV15, 
which is listed as a possible sgB[e] star in Table 5 based on its color and brightness, 
has been included in this plot as its location in Figure 10 is close to the area 
containing E-AGB stars, opening the possibility that it might be an LPV. 
A distance modulus of $27.67 \pm 0.09$ is found if all LPVs and DV15 are considered. 
The dispersion in the relation is $\pm 0.44$ magnitudes, which is 
larger than that in the Goldman et al. (2019) calibrating relation.
In any event, it is clear from the bottom panel of Figure 12 that the 
slope of the IC 2574 PLR differs from that found by Goldman et al. (2019), in 
the sense that it is shallower. If the distance modulus is 
based only on objects with periods in excess of 500 days, then a distance modulus of 
$27.78 \pm 0.09$ is found, with a dispersion of $\pm 0.37$ magnitudes. 
This distance modulus is in better agreement with the distance modulus listed 
in Table 1. The PLR based on this distance modulus is shown in Figure 12, and 
it can be seen from the bottom panel of Figure 12 that 
there is still a marked tendency for stars with periods of 550 days to fall above the 
expected PLR. The location of DV10 on the PLR suggests that it is an overtone pulsator.

	The result of placing the Goldman et al. (2019) relation on the IC 2574 
period $vs$ luminosity plane using the distance modulus listed in Table 1 is 
indicated with the red line in Figure 12. This relation passes through 
many of the longer period (and hence brighter) LPVs and possible LPVs in IC 2574, 
including DV15. The LPVs with periods $\leq 650$ days tend to 
fall well above the fitted relation. Some of the shorter period LPVs fall close to 
Sequence C', which is shown with a red dashed line. This supports the notion that 
some of the shorter period systems we have found in IC 2574 are likely overtone pulsators.

	The flat slope of the apparent PLRs in IC 2574 and NGC 2366 is due to a 
selection effect. There is a well-defined faint limit near [3.6] $\sim 17.7$ 
in the middle and bottom panels of Figure 12, which was imposed in the 
variable star search to ensure that full phase coverage could be obtained for 
any detected variable. This faint limit means that only the brightest fundamental mode 
pulsators with periods $\leq 500 - 650$ days are detected in the NGC 2366 and IC 2574 
observations, along with overtone pulsators. This biases the slope of the observed PLR 
to become flatter than the intrinsic relation.

\section{DISCUSSION AND CONCLUSIONS}

	Archival {\it Spitzer} IRAC observations have been used to identify and characterize 
bright LPVs and suspected evolved massive stars in the Magellanic-type galaxies Ho II, 
NGC 2366, and IC 2574. With distances of a few Mpc, these are among the nearest 
star forming dwarf galaxies outside of the Local Group, and 
so serve as stepping stones for understanding the stellar contents and SFHs of 
similar but more distant systems. Aside from 
having Magellanic-like morphologies, the three galaxies 
share other properties as well. As members of the M81 group they are more-or-less 
equidistant. They also have similar integrated brightnesses, and so likely have 
similar sub-solar metallicities. Finally, previous studies of stellar content in 
these galaxies suggest they have had similar, active SFHs during the past few hundred Myr, 
which is the timescale during which the progenitors of the variable stars studied here 
would have formed.

\subsection{LPVs and AGB stars}

	Luminous LPVs have been detected in all three galaxies. 
The locations of the LPVs on the $([4.5],[3.6]-[4.5])$ CMDs 
are consistent with them being near the bring end of the magnitude range 
occupied by E-AGB stars. The highly evolved nature of these objects is evident not only 
in their intrinsic brightnesses but also their [3.6]--[4.5] colors, which for many are 
indicative of an SED that has a significant contribution from warm circumstellar 
dust. Such emission is a signature of substantial mass loss, as would be expected if 
these objects are nearing the end of evolution on the AGB. Some of these LPVs 
are likely C stars, given that the galaxies have metallicities that are similar to 
those of the Magellanic Clouds, where C stars account for 
more than one quarter of the AGB component (e.g. Blum et al. 2006).

	The periods of the LPVs identified in the IRAC images range 
from 400 to 1300 days. LPVs with periods outside of 
this interval are missed in these data due to various 
biases. At short periods the detection of LPVs is compromised by photon noise and 
crowding, and this is evident in the truncated magnitude distributions of the 
PLRs of NGC 2366 and IC 2574 in Figure 12. The temporal sampling of the observations 
likely also complicates efforts to identify very short period variables. 
As for LPVs with very long periods, their detection is complicated by the cadence of the 
observations and stochastic effects due to the rapid evolutionary timescales of the 
most massive AGB stars. 

	The period distributions of LPVs in NGC 2366 and IC 2574 are very 
similar. While the temporal sampling of a suite of observations introduces obvious 
complications when comparing period distributions, this is mitigated to some extent 
by the similar sampling cadence of the light curves obtained from the 
SPIRITS images (i.e. ten observations per galaxy recorded at more-or-less similar 
intervals over a $\sim 1000$ day baseline). The agreement between period 
distributions includes suspected overtone pulsators (Section 6) that occur 
at periods $\leq 600$ days. To the extent that the period of an LPV and its peak 
brightness are loosely related to the mass of its progenitor (e.g. Takeuti et al. 
2013), then the similarity of the period distributions of NGC 2366 
and IC 2574 is broadly consistent with the SFHs of the galaxies in the age range 
over which the LPV progenitors formed having also been similar. The modest number 
of LPVs detected in Ho II complicates efforts to make meaningful comparisons with 
the period distributions of the other two galaxies, although there appears to be 
a deficiency in LPVs with periods $< 600$ days when compared with the others. 

	The spatial distributions of LPVs in all three galaxies differ from those of 
non-variable stars. In particular, the LPVs tend to be more diffusely 
distributed on the sky than the non-variable stars, the vast 
majority of which are likely also evolving on the AGB. The LPVs found 
in the IRAC images are among the brightest AGB stars in these galaxies, whereas the 
non-variable stars span a broader range of brightnesses, and hence ages, than 
the LPVs. The spatial distribution of the LPVs is then consistent with recent star 
formation being more diffusely distributed in these galaxies than the 
main body of older stars.

	There are holes in the distribution of bright AGB stars in Ho II. 
The absence of luminous AGB stars in these holes suggests that their interiors 
did not form stars for an extended period of time within the past few hundred Myr. 
These holes are reminiscent of the bubbles that are seen in HI maps of that galaxy. 
However, with the possible exception of the central HI bubble in 
Ho II, the voids found here do not have counterparts in the HI distribution. 
The orientations of NGC 2366 and IC 2574 make it difficult to determine if 
similar gaps are present in the projected distribution of AGB stars in those galaxies.

\subsection{LPV PLRs}

	The PLRs of the three galaxies constructed from the IRAC observations have 
different slopes. The PLR of LPVs in the [3.6] filter appears to be well-defined in 
Ho II, with a dispersion that is smaller than that in the measurements used by Goldman 
et al. (2019) to construct the calibrating relation. In fact, the dispersion 
measured from the Ho II data is an upper limit to the intrinsic scatter in the PLR 
of that galaxy given that there are uncertainties in the estimated periods, 
coupled with the likely presence of (1) stars that are not fundamental mode 
pulsators, and (2) LPVs that have optically thick circumstellar envelopes. 
Still, the slope of the Goldman et al. (2019) PLR in the top panel of Figure 12 
provides a reasonable match to the LPVs in Ho II, 
and the modest dispersion in the PLR suggests that the 
preliminary period estimates do not contain large errors. 
The distance estimated for Ho II from the Goldman et al. (2019) [3.6] PLR 
is consistent with that found from the tip of the RGB to within a tenth of a magnitude.

	In contrast to Ho II, the PLRs of NGC 2366 and IC 2574 obtained from the 
IRAC images are flatter than the Goldman et al. (2019) relation. 
LPVs in NGC 2366 that have log(P - days) $< 2.7$ 
fall above the expected PLR for fundamental mode pulsators, 
and some of these are probably overtone pulsators. If the 
most obvious candidates for overtone pulsators are not considered then a distance modulus 
for NGC 2366 that is consistent with the RGB-tip distance modulus is found.
The apparent PLR of IC 2574 is similarly skewed at shorter periods. Indeed, 
if the RGB-tip distance modulus is adopted for IC 2574 then 
the LPVs in IC 2574 with periods in excess of $\sim 650$ 
days fall along a sequence that follows the Goldman et al. (2019) PLR,
while those at shorter periods fall above this relation.

	The flat PLRs of NGC 2366 and IC 2574 are not due exclusively 
to contamination from overtone pulsators, as there is a photometric selection 
effect at play. The cut-off near [3.6] $\sim 18$ in the PLRs of NGC 2366 
and IC 2574 creates a bias in favor of detecting the 
brightest fundamental mode pulsators at shorter periods. Hence, it can be anticipated 
that many of the variables with periods $< 650$ days in these galaxies 
are likely first overtone pulsators that happen to fall above the PLR; LPVs below 
the relation are not detected due to the photometric cut-off imposed to detect 
variables. As stated in the previous paragraph, 
there are likely also overtone pulsators present with 
these periods. The period distribution of Ho II is deficient in LPVs with 
periods $< 650$ days when compared with the other two galaxies, and 
so any bias is not as obvious in the Ho II PLR. We also 
note that the apparent PLRs of NGC 2366 and IC 2574 PLRs is 
similar to that defined by LPVs in IC 1613 in Figure 6 of Goldman et al. (2019), 
suggesting that similar selection effects are likely at work in the LPV sample 
of that galaxy.

\subsection{Massive Stars}

	All three galaxies contain luminous red sources that may be among the 
most highly evolved massive stars in these systems. 
The brightest of these fall above the bright envelope of E-AGB stars in the 
$([4.5],[3.6]-[4.5])$ CMD, with many to the 
right of the foreground star sequence on the CMD. The reddest of 
these sources are likely a mix of sgB[e] stars and LBVs, although 
only sgB[e] stars occupy a distinct sequence in the CMD, and there is overlap between 
LPVs and LBVs in the CMD. While a number of potential LBVs are identified based on 
their light curves, it is likely that some (or all) of these may 
not turn out to be {\it bona fide} LBVs. Spectroscopic follow-up and longer term 
photometric monitoring will be essential for establishing the nature of these objects.

	While the sample size is small, there are hints of 
galaxy-to-galaxy differences in the numbers of sgB[e] stars. 
IC 2574 has the largest number of these objects, and this is reflected in the 
well-defined sgB[e] sequence in the left hand panel of Figure 10. 
Differences in massive star content may extend to other types of luminous red stars. 
While foreround contamination confounds efforts to identify individual 
RSGs with the existing data, the number counts of bright stars in Ho II suggest that 
that galaxy may host a larger population of very luminous RSGs than the other 
two galaxies. Given that sgB[e] stars and the most luminous RSGs are highly 
evolved massive objects, these results hint at galaxy-to-galaxy differences in the 
SFHs within the past few Myr.

	There is a high binary fraction among massive stars 
(e.g. Sana et al. 2012), and close binary systems will not follow the same 
evolutionary tracks as single stars. It has been suggested that 
some of the most luminous young stars in a galaxy might actually be close binary 
systems, in which mass exchange and mixing could extend the lifetime of the system 
and create stars that then do not follow the same age $vs$ mass $vs$ luminosity relation
as single objects (e.g. Smith \& Tombleson 2015). Binarity also provides one 
means of producing the disk that is seen around sgB[e] stars (e.g. Clark et al. 2013). 

	If the LBVs and sgB[e] stars identified here are binary systems 
then the numbers of these objects might be expected to show only modest galaxy-to-galaxy 
scatter given the evidence discussed in Section 1
of similar SFRs over the past few tens of Myr. In addition, if the 
possible massive stars found here are the result of interactions 
within binary systems then they would be distributed over a larger part of the sky 
near the host galaxy than single massive stars and HII regions, given that 
they live longer than single stars, and so have had more time to move away from their 
places of birth. The candidate massive stars found in the IRAC images appear to follow 
a diffuse sky distribution that is similar to that of the LPVs in all three galaxies. 

	We close by noting that the spatial resolution offered 
by the JWST in the Virgo cluster (15 -- 20 Mpc distance) is 
comparable to that achieved by {\it Spitzer} in the galaxies 
examined here ($\sim 3$ Mpc distance). The even better angular resolution that 
will be achieved with diffraction-limited VLOTs in $K$ will allow LPVs to be resolved in 
galaxies with distances comparable to those in the Coma cluster (i.e. 100 Mpc).
As is the case in NGC 2366 and IC 2574, crowding will likely limit the 
detection of fundamental mode pulsators in such target galaxies to those with the 
longest periods, and efforts to find such objects will thus need a suitable 
temporal sampling strategy. Observations of this nature will also greatly expand the 
sample of young massive evolved objects. The chances of characterizing a SN progenitor 
a few years before it explodes will increase as the 
sample of such highly evolved massive stars is expanded.

\acknowledgements{It is a pleasure to thank the anonymous referee for comments and 
suggestions that improved the manuscript. This research 
has made use of the NASA/IPAC Infrared Science
Archive (https://doi.org/10.26131/irsa22), 
which is funded by the National Aeronautics and Space Administration and
operated by the California Institute of Technology.
This research also has made use of the NASA/IPAC Extragalactic Database
(NED) \newline (https://doi.org/10.26132/ned1), which is operated 
by the Jet Propulsion Laboratory, California Institute of
Technology, under contract with the National Aeronautics and Space Administration. 
The Pan-STARRS1 Surveys (PS1) and the PS1 public science archive have been made 
possible through contributions by the Institute for Astronomy, the University of 
Hawaii, the Pan-STARRS Project Office, the Max-Planck Society and its participating 
institutes, the Max Planck Institute for Astronomy, Heidelberg and the Max Planck 
Institute for Extraterrestrial Physics, Garching, The Johns Hopkins University, 
Durham University, the University of Edinburgh, the Queen's University Belfast, the 
Harvard-Smithsonian Center for Astrophysics, the Las Cumbres Observatory Global 
Telescope Network Incorporated, the National Central University of Taiwan, the Space 
Telescope Science Institute, the National Aeronautics and Space Administration under 
Grant No. NNX08AR22G issued through the Planetary Science Division of the NASA Science 
Mission Directorate, the National Science Foundation Grant No. AST-1238877, the 
University of Maryland, Eotvos Lorand University (ELTE), the Los Alamos National 
Laboratory, and the Gordon and Betty Moore Foundation.}

\end{document}